\documentclass[12pt,JPCM]{iopart}
\expandafter\let\csname equation*\endcsname\relax
\expandafter\let\csname endequation*\endcsname\relax
\usepackage{amssymb,amsmath,amsfonts,multicol,multirow,longtable,array,mathpazo}
\usepackage{lscape}
\usepackage[version=3]{mhchem} % Formula subscripts using \ce{}
\usepackage{appendix}
\usepackage{soul} %text highlighting
\usepackage[usenames]{color}
\usepackage{makecell}
\usepackage{enumerate}
\usepackage{xr}
\usepackage[T1]{fontenc} % Use modern font encodings
\usepackage{booktabs}
\usepackage{tikz}
\usepackage{threeparttable}
\usepackage{graphicx}
\usepackage{subfigure}
\usepackage{colortbl}
\usepackage{framed}
\usepackage{verbatim}
\usepackage{amsbsy}
\usepackage{bm}% bold math
\usepackage{bbm}
\usepackage[normalem]{ulem}
\usepackage{float}
\usepackage[linesnumbered,boxed]{algorithm2e}
\usepackage{algorithm2e}
\usepackage{mathrsfs}
\usepackage[utf8]{inputenc}
\usepackage[numbers,sort&compress]{natbib}

\newcounter{xscheme}

\newfloat{Algorithm}{htbp}{alg}
\floatname{Algorithm}{Algorithm}

\newcounter{exe}[figure]
\newcommand{\iexe}{\refstepcounter{exe}\the\value{exe}:}

\begin{document}

%\title[Author guidelines for IOP Publishing journals in  \LaTeXe]{How to prepare and submit an article for
%publication in an IOP Publishing journal using \LaTeXe}

\title{SOiCI and iCISO: Combining iterative configuration interaction with spin-orbit coupling in two ways}

\author{Ning Zhang}

\address{Beijing National Laboratory for Molecular Sciences, Institute of Theoretical and Computational Chemistry,
	College of Chemistry and Molecular Engineering, Peking University, Beijing 100871, China}
%\ead{submissions@iop.org}

\author{Yunlong Xiao}
\address{Beijing National Laboratory for Molecular Sciences, Institute of Theoretical and Computational Chemistry,
	College of Chemistry and Molecular Engineering, Peking University, Beijing 100871, China}

\author{Wenjian Liu}
\address{Qingdao Institute for Theoretical and Computational Sciences, Institute of Frontier and Interdisciplinary Science,
	Shandong University, Qingdao, Shandong 266237, China}
\ead{liuwj@sdu.edu.cn}

\vspace{10pt}

\begin{abstract}
The near-exact iCIPT2 approach for strongly correlated systems of electrons, which stems from the combination of iterative configuration interaction (iCI,
an exact solver of full CI) with configuration selection for static correlation
and second-order perturbation theory (PT2) for dynamic correlation, is extended to the relativistic domain.
In the spirit of spin separation, relativistic effects are treated in two steps: scalar relativity is treated by the infinite-order,
spin-free part of the exact two-component (X2C) relativistic Hamiltonian, whereas spin-orbit coupling (SOC) is treated by
the first-order, Douglas-Kroll-Hess-like SOC operator
derived from the same X2C Hamiltonian. Two possible combinations of iCIPT2 with SOC are considered, i.e., SOiCI and iCISO.
The former treats SOC and electron correlation on an equal footing, whereas the latter treats SOC in the spirit of state interaction, by constructing and diagonalizing
an effective spin-orbit Hamiltonian matrix in a small number of correlated scalar states. Both double group and time reversal symmetries
are incorporated to simplify the computation. Pilot applications reveal that SOiCI is very accurate for the spin-orbit splitting (SOS) of heavy atoms,
whereas the computationally very cheap iCISO can safely be applied to the SOS of light atoms and even of
systems containing heavy atoms when SOC is largely quenched by ligand fields.

\end{abstract}

\section{Introduction}

% Motivation

It has long been recognized\cite{Pyykko1988,pyykko2012,Autschbach2012} that both relativistic and correlation effects ought to be accounted for in accurate quantum mechanical descriptions of
the electronic structure not only of systems containing heavy elements (say, $Z>30$) but also of systems composed only of light elements ($Z\le 30$).
Literally, relativity is in the Hamiltonian, whereas correlation is in the wave function parameterized by a particular ansatz\cite{LiuMP}.
As the advent of the continuous and complete ``Hamiltonian ladder''\cite{LiuPhysRep,IJQCrelH}, the relativity problem in quantum chemistry can be
regarded as solved, in the sense that one can just pick up the right Hamiltonian according to the target physics/chemistry and accuracy. In particular,
all relativistic Hamiltonians\cite{LiuMP,Hess2000,Saue2011,Peng2012,X2C2016,X2CBook2017,X2CSOCBook2017}, including
the effective QED\cite{LiuPhysRep,IJQCrelH,np-eQED,Kutzelnigg2012,eQED,PCCPNES,IJQCeQED,eQEDBook2017,LiuPerspective2020,LiuSciChina2020}, can be written in the same second-quantized form
\begin{equation}
H=E_{\mathrm{ref}}+ f_p^q\{a^p_q\}+ \frac{1}{2}g_{pq}^{rs}\{a^{pq}_{rs}\},\quad f_p^q=\langle \psi_p|\hat{f}|\psi_q\rangle,\quad g_{pq}^{rs}=(\psi_p\psi_r|g(1,2)|\psi_q\psi_s),\label{Hop}
\end{equation}
where $E_{\mathrm{ref}}$ is the expectation value of the Hamiltonian $H$ over a reference state $|0\rangle$ (built up with
an orthonormal set of one-particle functions $\{\psi_p\}$), with respect to which the normal ordering
of the one-body ($a^p_q=a_p^\dag a_q$) and two-body ($a^{pq}_{rs}=a_p^\dag a_q^\dag a_s a_r$) excitation operators has been taken. Different Hamiltonians
differ only in the effective one-electron ($f_p^q$) and two-electron ($g_{pq}^{rs}$) integrals. As such, under the no-pair approximation, any relativistic Hamiltonian can directly be combined with all orbital-product-based
wave function methods\cite{npCorrelation2016}. It is just that the breaking of spin symmetry and the concurrent appearance of complex algebra in the presence
of spin-orbit coupling (SOC) render not only the implementation difficult but also the computation expensive.
Nevertheless, many sophisticated relativistic correlated wave function methods have been made available for use, including
four- (4C) or two-component (2C) many-body perturbation theory\cite{4C-MP21994,2C-MRPT22014}, coupled-cluster\cite{4C-CCSD1995,4C-CCSD1996,2C-CC2001,2C-CC2005,2C-CC2007,4C-IHFSCC2000,4C-IHFSCC2001,4C-FSCC2001,4C-CC2007,4C-CC2010,2C4C-CC2011,4C-CC2016,2C-CC2017,Cheng2C-CC2018,Saue4C-CC2016,Saue4C-CC2018,2C-EOM-CCSD2019,Cheng2C-CCrev,SOC-CCSDT-GPU2021,CVS-EOM-CCSD2021},
configuration interaction (CI)\cite{4C-CISD1993,2C-CI1996,2C-CI2001,4C-GASCI2003,4C-CI2008,2C-CICC2012,Fleig2012,2C-MRCI2020}, multiconfiguration self-consistent field
\cite{RASSOC,CASSI1986,CASSI1989,4C-MCSCF1996,2C-CASSCF1996,2C-CASSCF2003,4C-CI-MCSCF2006,2C-CASSCF2013,4C-MCSCF2008,4C-CASPT22008,4C-CASSCF2015,4C-CASSCF2018,4C-icMRCI2015,RelDMRG2005,RelDMRG2014,4C-MR2018,LixiaosongX2CCASSCF},
density-matrix renormalization group\cite{4C-DMRG2014,4C-DMRG2018,4C-DMRG2020,MPSSI2021}, and full configuration interaction quantum Monte Carlo\cite{4c-FCIQMC}.
It should be clear from the outset that, under the no-pair approximation, a 4C wave function method
is computationally identical with its 2C counterpart after integral transformations, whether the Hamiltonian is all-electron or valence-only.
Even the 4C integral transformations can be made identical with the 2C ones if the quasi-4-component (Q4C) relativistic Hamiltonian\cite{Q4C,X2C2007} is adopted, which
does not suffer from picture change errors\cite{PCE} that otherwise plague all 2C relativistic Hamiltonians.
Note also that the correlation contribution of negative energy states can readily be accounted for in both the 4C and 2C frameworks, so as to go beyond the no-pair approximation
\cite{LiuPhysRep,IJQCrelH,Kutzelnigg2012,PCCPNES,IJQCeQED,eQEDBook2017,LiuPerspective2020,LiuSciChina2020}.
As such, it is merely a matter of taste to work with 4C or 2C approaches. Such methods are imperative for core properties of heavy elements
or valence properties involving $np$ ($n>5$) orbitals. However, the situation is different for most chemical systems where SOC is
not very strong, such that the $jj$ (or $\omega\omega$) coupling scheme underlying 4C/2C approaches becomes inappropriate.
In such cases, it is more appropriate to work with the intermediate coupling scheme\cite{bagus2015intermediate}, where $LS$-coupled many-electron basis functions
are allowed to interact via SOC. In other words, the SOC is postponed to the correlation step, such that
the molecular orbitals (MO) and hence the integrals can be chosen to be real-valued. For this reason, such approaches are usually called one-component (1C).
The interplay between SOC and electron correlation
can be accounted for in two ways, one-step or two-step\cite{Marian2001,marian2012SOCISC}. The former type of approaches\cite{DGCIa,SpinUGA1,DGCIb,DGCIc,DetSOCI1997,CI-SOC1998,SPOCK2006,WF-CC2008,WF-CC2011,CAS-SOC2013,sf-X2C-EOM-SOC2017,HBCISOC2017}
aims to treat spin-orbit and electron-electron interactions on an equal footing, whereas the latter type of approaches
\cite{Hess1982,CIPSO1983,CI-SOC1997,CI-SOC1998,SOCsingle-1,SOCsingle-2,SOCsingle-3,MRCI-SOC2000,Teichteil2000,DFTMRCI-SOC2002,CASPT2-SOC2004,SI-SOC2006,EOMIPSO2008,mai2014perturbational,DMRGSO2015,DMRGSO2016,nosiDMRGSO2016,cheng2018perturbative,ChengEOMCCSO2020,WangMP2020,Suo-SOC2021,vanWullen2021}
amounts to treating SOC after correlation, by
constructing and diagonalizing an effective spin-orbit Hamiltonian matrix over a set of close-lying correlated scalar states.

In this work, we extend the recently proposed iCIPT2 approach\cite{iCIPT2,iCIPT2New} to the relativistic domain within the 1C framework. iCIPT2 stems from
the combination of iterative configuration interaction (iCI, an exact solver of full CI)\cite{iCI} with configuration selection for static correlation
and second-order perturbation theory for dynamic correlation, and belongs to the `modern family' of selected CI methods\cite{LambdaCI2014,ACI2016,ACI2017,ACI2018,ASCI2016,ASCI2018PT2,ASCI2020,HBCI2016,HBCI2017a,HBCI2017b,HBCI2017c,SHBCI2018,HBCI-Cr2,HBCI2020,DressedSCI2018,CD-FCI,CIPSI-DMCd,CIPSI-DMCe}
for strongly correlated systems of electrons. In particular, the use of configuration state functions (CSF) as the many-electron basis
and the tabulated unitary group approach (TUGA)\cite{iCIPT2} for fast evaluation and reuse of the basic coupling coefficients between CSFs
allows an easy extension of iCIPT2 to the treatment of SOC in two ways, one-step SOiCI and two-step iCISO, which are to be discussed in detail in  Sec. \ref{SecSOiCI},
after introducing the spin-separated exact two-component (X2C) relativistic Hamiltonian\cite{X2CSOCBook2017,X2CSOC1,X2CSOC2} in Sec. \ref{SecsfX2C} and symmetry adaptation of
the many-electron basis in Sec. \ref{doublegroup}.
Pilot applications are then presented in Sec. \ref{Results}, followed by concluding remarks in Sec. \ref{Conclusion}.

The following notations are to be used throughout. A configuration of in total $n$ ordered spatial orbitals $\{\phi_r\}$ with occupation numbers $\{n_r\}$ is denoted as $|I\rangle$,
which can generate
\begin{eqnarray}
N_{S}^{\mathrm{c}}=\frac{2S+1}{S_{\mathrm{high}}+S+1}C_{N_{\mathrm{o}}}^{S_{\mathrm{high}}-S},\quad S_{\mathrm{high}}=\frac{1}{2}N_{\mathrm{o}}\label{Ncsf}
\end{eqnarray}
CSFs of spin $S$ if it has $N_o$ singly occupied orbitals. Since each CSF
is characterized uniquely by a vector $\mathbf{d}$ of the Shavitt step numbers $\{d_r\}$ ($d_r=0$ if orbital $r$ is not occupied; $d_r=1$ if orbital $r$ is singly occupied
and spin-up coupled with orbital $r-1$; $d_r=2$ if orbital $r$ is singly occupied
and spin-down coupled with orbital  $r-1$; $d_r=3$ if orbital $r$ is doubly occupied)\cite{Shavitt1977}, a CSF can be denoted simply as $|I\mathbf{d}\rangle$.
When necessary, the notation will be expanded to $|I\mathbf{d}S\rangle$, even though the spin $S$ can be derived automatically from $\mathbf{d}$.
The components of $|I\mathbf{d}\rangle$ are denoted accordingly as $|I\mathbf{d}M\rangle$ (or $|I\mathbf{d}SM\rangle$) with $M \in [-S, S]$.
After double group and time reversal symmetry adaption (see Sec. \ref{doublegroup} and Appendices \ref{AppDG} and \ref{AppTR}),
the Kramers paired components of $|I\mathbf{d}\rangle$ will be denoted
as $|I\mathbf{d}\mu\rangle$ (or $|I\mathbf{d}S\mu\rangle$) with $\mu\in[1,n_{\gamma}^{\mathbf{d}}]$, where $n_{\gamma}^{\mathbf{d}}$
is the dimension of an irreducible representation (irrep) $\gamma$ of a double group.

\section{Two-step Relativistic Hamiltonian: sf-X2C+so-DKH1}\label{SecsfX2C}
A two-step relativistic Hamiltonian is needed to postpone the treatment of SOC to the correlation step.
It has been shown\cite{X2CSOCBook2017,X2CSOC1,X2CSOC2} that the X2C Hamiltonian\cite{X2C2005,X2C2009}, albeit defined only algebraically,
can be separated into a spin-free  part (sf-X2C) and a family of spin-dependent operators. Among the latter, the
first-order Douglas-Kroll-Hess-like spin-orbit operator (so-DKH1) is the simplest variant. Without going into further details,
the sf-X2C+so-DKH1 Hamiltonian reads
\begin{eqnarray}
H&=&H_{sf}+H_{so},\label{approximateH}\\
H_{sf}&=&\sum_{pq}[h_{+,sf}^{\mathrm{X2C}}]_{pq}E_{pq}+\frac{1}{2} \sum_{pqrs}(pq|rs) e_{pq,rs},\label{TCHsf}\\
H_{so}&=&\mathbbm{i} \sum_{pq}\sum_{l\in {x,y,z}} [H_{so}^l]_{pq} T_{pq}^l,\quad \mathbf{H}_{so}^l=\mathbf{h}_{SO,1e}^l+\mathbf{f}_{SO,2e}^l,\label{SOCoper}\\
E_{pq}&=&\sum_{\sigma}a_{p\sigma}^\dag a_{q\sigma},\quad e_{pq,rs}=E_{pq}E_{rs}-\delta_{qr}E_{ps},\\
T_{pq}^x&=&a_{p\alpha}^\dag a_{q\beta}+ a_{p\beta}^\dag a_{q\alpha},\quad T_{pq}^y=-\mathbbm{i} (a_{p\alpha}^\dag a_{q\beta}- a_{p\beta}^\dag a_{q\alpha}),\nonumber\\
T_{pq}^z&=&a_{p\alpha}^\dag a_{q\alpha}- a_{p\beta}^\dag a_{q\beta},
\end{eqnarray}
where $p,q,r,s$ refer to real-valued MOs and the two-electron integrals have been written in the Mulliken notation.
The one-electron term $\mathbf{h}_{+,sf}^{\mathrm{X2C}}$ in $H_{sf}$ \eqref{TCHsf} is obtained simply by block-diagonalizing the
spin-free part of the one-electron Dirac equation represented in a kinetically balanced basis\cite{RKB},
\begin{eqnarray}
\mathbf{h}^{\mathrm{X2C}}_{+,sf}&=&\mathbf{R}_{+,0}^\dagger(\mathbf{V}_{ne}+\mathbf{T}\mathbf{X}_0+\mathbf{X}_0^\dagger\mathbf{T}
+\mathbf{X}_0^\dagger[\frac{\alpha^2}{4}\mathbf{W}_{sf}-\mathbf{T}]\mathbf{X}_0)\mathbf{R}_{+,0},\\
\mathbf{R}_{+,0}&=&(\mathbf{S}^{-1}\tilde{\mathbf{S}}_{+,0})^{-\frac{1}{2}}
=\mathbf{S}^{-\frac{1}{2}}(\mathbf{S}^{-\frac{1}{2}}\tilde{\mathbf{S}}_{+,0}\mathbf{S}^{-\frac{1}{2}})^{-\frac{1}{2}}\mathbf{S}^{\frac{1}{2}},\label{Rmat}\\
\tilde{\mathbf{S}}_{+,0}&=&\mathbf{S}+\frac{\alpha^2}{2}\mathbf{X}^{\dagger}_{0}\mathbf{TX}_{0},\\
\mathbf{X}_0&=&\mathbf{B}_{+}\mathbf{A}^{-1}_{+}=\mathbf{B}_{+}\mathbf{A}_{+}^{\dagger}(\mathbf{A}_{+}\mathbf{A}_{+}^{\dagger})^{-1},\label{Xmat}
\end{eqnarray}
where $\alpha$ is the fine-structure constant, $\mathbf{S}$,
$\mathbf{T}$ and $\mathbf{V}_{ne}$ are the respective  matrices of the nonrelativistic
metric, kinetic energy and nuclear attraction ($V_{ne}$), whereas
$\mathbf{W}_{sf}$ is the matrix of the ``small-component potential''
$W_{sf}=\vec{p}\cdot V_{ne}\vec{p}$. The decoupling matrix $\mathbf{X}_0$ \eqref{Xmat} is
simply the ratio between the coefficients of the small ($\mathbf{B}_+$) and large ($\mathbf{A}_+$)
components of the scalar, two-component Dirac orbitals of positive energies. Note that scalar two-electron picture change corrections
have been neglected here since this is known to be a very good approximation for valence properties\cite{1eSR2005}.
The so-DKH1 operator $H_{so}$ \eqref{SOCoper} is composed of a
one-electron term $\mathbf{h}_{SO,1e}$\cite{X2CSOC1} and a mean-field approximation  $\mathbf{f}_{SO,2e}$\cite{X2CSOC2} to the two-electron
spin-orbit interaction, viz.,
\begin{eqnarray}
\mathbf{h}_{SO,1e}^l&=&\frac{\alpha^2}{4}\mathbf{R}_{+,0}^{\dagger}\mathbf{X}_0^\dagger\mathbf{w}^l\mathbf{X}_0\mathbf{R}_{+,0},\quad l\in\{x,y,z\},\label{hSO}\\
\mathbf{f}_{SO,2e}^l&=&\frac{\alpha^2}{4}\mathbf{R}_{+,0}^{\dagger}[\mathbf{g}^{LL,l}+\mathbf{g}^{LS,l}\mathbf{X}_{0}+
\mathbf{X}^{\dagger}_{0}\mathbf{g}^{SL,l}+\mathbf{X}^{\dagger}_{0}\mathbf{g}^{SS,l}\mathbf{X}_{0}]\mathbf{R}_{+,0}.\label{FockSO}
\end{eqnarray}
Here, $\mathbf{w}^l$ is the matrix of
$w^l= (\vec{p}V_{ne}\times\vec{p})_l=\varepsilon_{lmn}p_m V_{ne} p_n$ in the basis of atomic orbitals (AO), viz.,
\begin{eqnarray}
w^l_{\mu\nu}&=&\varepsilon_{lmn}\langle\mu_m|V_{ne}\nu_n\rangle=-w^l_{\nu\mu},\quad \mu_m=\partial_m \mu,\label{wlmat}
\end{eqnarray}
where $\varepsilon_{lmn}$ is the Levi-Civita symbol. The effective one-body spin-orbit integrals
$\mathbf{g}^{XY,l}$ ($X, Y \in\{ L, S \}$) are defined as\cite{X2CSOC2}
\begin{eqnarray}
g^{LL,l}_{\mu\nu}&=&-2\sum_{\lambda\kappa}K^{l}_{\lambda\mu,\kappa\nu}P^{SS}_{\lambda\kappa}=-g^{LL,l}_{\nu\mu},\quad l\in\{x,y,z\},\label{gll}\\
g^{LS,l}_{\mu\nu}&=&-\sum_{\lambda\kappa}(K^l_{\mu\lambda,\kappa\nu}+K^l_{\lambda\mu,\kappa\nu})P_{\lambda\kappa}^{LS}=-g^{SL,l}_{\nu\mu},\label{gls}\\
g^{SL,l}_{\mu\nu}&=& \sum_{\lambda\kappa}(K^l_{\mu\lambda,\kappa\nu}+K^l_{\mu\lambda,\nu\kappa})P_{\lambda\kappa}^{SL}=-g^{LS,l}_{\nu\mu},\label{gsl}\\
g^{SS,l}_{\mu\nu}&=&-2\sum_{\lambda\kappa}(K^l_{\mu\nu,\kappa\lambda}+K^l_{\mu\nu,\lambda\kappa}-K^l_{\mu\lambda,\nu\kappa})
P^{LL}_{\lambda\kappa}=-g^{SS,l}_{\nu\mu},\label{gss}\\
K^{l}_{\mu\nu,\kappa\lambda}&=&\sum_{mn}\varepsilon_{lmn}(\mu_m\nu|\kappa_n\lambda)=-K^{l}_{\kappa\lambda,\mu\nu},\nonumber\\
\mu_m&=&\partial_m\mu,\quad l,m,n\in\{x,y,z\},\label{SOintegrals}\\
\mathbf{P}^{LL}&=&\mathbf{R}_{+,0}\mathbf{P}\mathbf{R}_{+,0}^{\dagger},\quad
\mathbf{P}^{LS}=\mathbf{P}^{LL}\mathbf{X}^{\dagger}_0=\mathbf{P}^{SL\dag},\quad\mathbf{P}^{SS}=\mathbf{X}_{0}\mathbf{P}^{LL}\mathbf{X}^{\dagger}_0.
\end{eqnarray}
Here,
$\mathbf{P}=\frac{1}{2}(\mathbf{P}^\alpha+\mathbf{P}^\beta)$ is the real-valued, spin-averaged molecular density matrix,
with $\mathbf{P}^\alpha$ and $\mathbf{P}^\beta$ being the converged sf-X2C-ROHF (restricted open-shell Hartree-Fock) spin density matrices.
The terms \eqref{gls} and \eqref{gsl} as well as the first two terms in Eq. \eqref{gss} arise from the Coulomb-exchange interaction and represent
the so-called spin-same-orbit coupling, whereas
the term \eqref{gll} and the third term of Eq. \eqref{gss} originate from
the Gaunt-exchange interaction and hence represent the spin-other-orbit coupling. The Gaunt-direct interaction
vanishes due to spin averaging. Note in passing that
$\mathbf{w}^l$ \eqref{wlmat} and $\mathbf{g}^{XY,l}$ \eqref{gll}-\eqref{gss} are all real-valued and antisymmetric.
%In view of the short-range nature of spin-orbit interactions, a one-centre approximation to
%the integrals $K^{l}_{\mu\nu,\kappa\lambda}$ \eqref{SOintegrals} is further invoked in the calculations. In this case, only the atomic blocks of the
%molecular density matrix $\mathbf{P}^{XY}$ contribute to $\mathbf{g}^{XY,l}$. Yet, $\mathbf{f}_{SO,2e}^l$ is still a full matrix.
It deserves to be mentioned that $H_{so}$ \eqref{SOCoper} will reduce\cite{X2CSOC2} to
the mean-field Breit-Pauli spin-orbit Hamiltonian or the original mean-field so-DKH1\cite{SOMF1,SOMF2}
if both the decoupling matrix $\mathbf{X}_{0}$ \eqref{Xmat} and the renormalization matrix $\mathbf{R}_{+,0}$ \eqref{Rmat} are set to identity or the free-particle counterparts
in $\mathbf{h}_{SO,1e}^l$ \eqref{hSO} and $\mathbf{f}_{SO,2e}^l$ \eqref{FockSO}.
It is hence clear that so-DKH1 gains accuracy withiout overhead, as demonstrated before\cite{sf-X2C-EOM-SOC2017,HBCISOC2017,X2CSOC1,XTDDFT-SOC}.

To facilitate the use of the spin-dependent unitary group approach (UGA)\cite{SpinUGA1, SpinUGA2} for SOC, we further convert the so-DKH1 operator $H_{so}$ \eqref{SOCoper} from Cartesian to tensor form
\begin{eqnarray}
H_{so}       &=&\sum_{pq} [H_{so}^1]_{pq}T_{pq}^{1,-1}+[H_{so}^0]_{pq}T_{pq}^{1,0}+[H_{so}^{-1}]_{pq}T_{pq}^{1,1}, \label{TensorHso}\\
\mathbf{H}_{so}^1 & =& \mathbbm{i} \mathbf{H}_{so}^{x} - \mathbf{H}_{so}^{y},\quad
\mathbf{H}_{so}^0= \mathbbm{i}\sqrt{2}\mathbf{H}_{so}^{z}, \quad
\mathbf{H}_{so}^{-1}= -\mathbbm{i}\mathbf{H}_{so}^{x} - \mathbf{H}_{so}^{y},\\
T_{pq}^{1,-1} & =& a_{p\beta}^\dagger a_{q\alpha}, \quad
T_{pq}^{1, 0} = \frac{1}{\sqrt{2}}(a_{p\alpha}^\dagger a_{q\alpha}-a_{p\beta}^\dagger a_{q\beta}), \quad
T_{pq}^{1, 1} = -a_{p\alpha}^\dagger a_{q\beta}.
\end{eqnarray}
The matrix elements of the rank-1 tensor operator $T_{pq}^{1,\gamma}$ over $\{|I\mathbf{d}'S'M'\rangle\}_{M'=-S'}^{S'}$ and $\{|J\mathbf{d}SM\rangle\}_{M=-S}^S$ then read
\begin{equation}
\begin{split}
\langle I\mathbf{d}' S'M'|T^{1,\gamma}_{pq}|J\mathbf{d}SM\rangle
&  = \frac{(-1)^{S'-M'+S_{n+1}+S-\frac{1}{2}}}{\sqrt{3}}
\begin{pmatrix}
S'  & 1      & S \\
-M' & \gamma & M \\
\end{pmatrix}
\left\{
\begin{array}{ccc}
S'          & S           & 1       \\
\frac{1}{2} & \frac{1}{2} & S_{n+1} \\
\end{array}
\right\}^{-1} \\
& \langle I\mathbf{d}'_{n+1} S_{n+1}M_{n+1} | e_{p,n+1;n+1,q}+\frac{1}{2}E_{pq}|J\mathbf{d}_{n+1}S_{n+1}M_{n+1}\rangle,
\end{split} \label{SpinUGA}
\end{equation}
where $|I\mathbf{d}_{n+1}S_{n+1}\rangle$ refers to a CSF with $n+1$ spatial orbitals, which can be characterized by padding the step number $d_{n+1}$ to the end of $\mathbf{d}$.
This means that such matrix elements can be calculated in terms of the $U(n+1)$ generators, by only slight modifications of
the current implementation of TUGA\cite{iCIPT2}.
The step numbers $d_{n+1}'$ and $d_{n+1}$ as well as the spin $S_{n+1}$ are determined by $S'$ and $S$ as follows:
\begin{equation}
\left\{
\begin{array}{cccc}
S_{n+1} =  S+\frac{1}{2}, & d_{n+1}'=1, & d_{n+1} =1, & \text{if }S'=S,   \\
S_{n+1} =  S-\frac{1}{2}, & d_{n+1}'=2, & d_{n+1} =2, & \text{if }S'=S,   \\
S_{n+1} =  S+\frac{1}{2}, & d_{n+1}'=2, & d_{n+1} =1, & \text{if }S'=S+1, \\
S_{n+1} =  S-\frac{1}{2}, & d_{n+1}'=1, & d_{n+1} =2, & \text{if }S'=S-1. \\
\end{array}
\right.\label{twoSval}
\end{equation}
For the case of $S'=S$, we adopt here $S_{n+1}=S+\frac{1}{2}$.
In view of the Wigner-Eckart theorem, the left hand side of Eq. \eqref{SpinUGA} can further be simplified to
\begin{equation}
\langle I\mathbf{d}' S'M'|T^{1,\gamma}_{pq}|J\mathbf{d}SM\rangle = (-1)^{S'-M'}
\begin{pmatrix}
S'  & 1      & S \\
-M' & \gamma & M \\
\end{pmatrix}
\langle I\mathbf{d}' S'||T^{1}_{pq}||J\mathbf{d}S\rangle, \label{Hsd_MatElmt}
\end{equation}
where the reduced matrix elements can be calculated as
\begin{equation}
\begin{split}
\langle I\mathbf{d}' S'||T^{1}_{pq}||J\mathbf{d}S\rangle
&  = \frac{(-1)^{S_{n+1}+S-\frac{1}{2}}}{\sqrt{3}}
\left\{
\begin{array}{ccc}
S'          & S           & 1       \\
\frac{1}{2} & \frac{1}{2} & S_{n+1} \\
\end{array}
\right\}^{-1} \\
& \langle I\mathbf{d}'_{n+1} S_{n+1}S_{n+1} | e_{p,n+1;n+1,q}+\frac{1}{2}E_{pq}|J\mathbf{d}_{n+1}S_{n+1}S_{n+1}\rangle.\label{RME}
\end{split}
\end{equation}

\section{Double Group and Time Reversal Symmetries}\label{doublegroup}
Unlike the spin-free case, the spin and spacial degrees of freedom are coupled in the presence of SOC. In particular,
for a system of odd number of electrons, the eigenfunctions of the Hamiltonian \eqref{approximateH} have a half-integral spin and hence
change sign under a rotation $2\pi$ about an arbitrary axis. Such an operation (denoted as $\bar{E}$) commutes with all symmetry
operations and can hence be added into a single group $G$, thereby doubling the order $|G|$ of the group. The so-obtained
group $G^\ast$ is hence called double group, even though the number of irreps is not doubled (the original and extra irreps are called boson and fermion irreps, respectively).
Consider $D_{2h}^\ast$, which is a non-Abelian subgroup of $SU(2)\times C_i$, with $C_i=\{E,i\}$ being the inversion point group.
Since the elements of $SU(2)$ can be parameterized in the spin $(\alpha,\beta)$ basis as
\begin{eqnarray}
\mathbf{R}^{\frac{1}{2}}(\vec{\boldsymbol{n}},\varphi)&=&\exp\left(-\frac{\mathbbm{i}}{2}\varphi\vec{\boldsymbol{\sigma}}\cdot\vec{\boldsymbol{n}}\right),\quad \varphi\in[0,4\pi)\\
&=&\mathbf{I}_2\cos\frac{\varphi}{2}-\mathbbm{i}(\vec{\boldsymbol{\sigma}}\cdot\vec{\boldsymbol{n}})\sin\frac{\varphi}{2},\label{SU2def}
\end{eqnarray}
with $\vec{\boldsymbol{\sigma}}$ and $\mathbf{I}_2$ being the vector of Pauli spin matrices and two-dimensional unit matrix, respectively,
it is easy to see that every element $g$ of $D_{2h}^\ast$ can be factorized as
\begin{eqnarray}
g=abc,\quad &a&\in A=\{\mathbf{I}_2,-\mathbbm{i}\sigma_x,-\mathbbm{i}\sigma_y,-\mathbbm{i}\sigma_z\},\nonumber\\
&b&\in B=\{\mathbf{I}_2,-\mathbf{I}_2\},\quad c\in C_i=\{E,i\},\label{ABC}
\end{eqnarray}
where $-\mathbbm{i}\sigma_{\mu}$ is just $\mathbf{R}^{\frac{1}{2}}(\mu,\pi)$ with $\mu=x, y, z$, whereas
 $-\mathbf{I}_2$ in group $B$ is just $\bar{E}$ for it is equal to $\mathbf{R}^{\frac{1}{2}}(\vec{\boldsymbol{n}},2\pi)$
  regardless of the rotation axis $\vec{\boldsymbol{n}}$.
By definition, the inversion $i$ acts trivially on spin coordinates. Likewise, the elements in $B$ acts trivially on spatial coordinates.
In contrast, the elements in $A$ act on both spin and spatial coordinates and should be reinterpreted as operations
$E, C_{2x}, C_{2y}, C_{2z}$, respectively, when acting on
spatial coordinates. That is, in the absence of $B$, we will have $A=D_2$ and hence $A\times C_i=D_{2h}$.
Since any element $g$ of $D_{2h}^\ast$ (and its subgroups) does not mix CSFs of different spins or of the same spin but different Shavitt step vectors,
it is necessary to consider only the action of $g$ on the components $\{|J\mathbf{d}M\rangle\}$
\begin{equation}
	|J\mathbf{d}M\rangle = \mathcal{A}\left[ |\phi_1\rangle \otimes |\phi_2\rangle \otimes \cdots \otimes |\phi_N\rangle \otimes |SM\rangle\right],\quad M\in[-S,S]
\end{equation}
of a single CSF $|J\mathbf{d}\rangle$. Here, the spatial orbitals have been adapted to the 1D irreps of $G$ (and hence $G^\ast$).
Since the antisymmetrizer $\mathcal{A}$ commutes with $g$, the action of $g$ takes the following form
\begin{eqnarray}
g|J\mathbf{d}M\rangle&=& \mathcal{A} \left[ (g|\phi_1\rangle)\otimes (g|\phi_2\rangle)\otimes\cdots\otimes (g|\phi_N\rangle)\otimes (g|SM\rangle)\right]\\
&=&\mathcal{A} \left[ (|\phi_1\rangle\otimes |\phi_2\rangle\otimes\cdots\otimes |\phi_N\rangle) \mathbf{V}^{(g)} \otimes \sum_{M^\prime} |SM^\prime\rangle \mathbf{S}^{(g)}_{M^\prime M}\right]\\
&=&\sum_{M^\prime}|J\mathbf{d}M^\prime\rangle \mathbf{U}^g_{M^\prime M}, \quad \mathbf{U}^{(g)}= \mathbf{V}^{(g)}\otimes\mathbf{S}^{(g)},\quad \mathbf{V}^g\in\{\pm 1\}. \label{VSmat}
\end{eqnarray}
The spin rotation matrices $\mathbf{S}^{(g)}$ are constructed explicitly in Appendices \ref{AppDG} and \ref{AppTR}.

The so-constructed symmetric many-electron basis functions $\{|J\mathbf{d}\mu\rangle\}$ can further be made to form Kramers pairs via time reversal symmetry (see Appendix \ref{AppTR}),
so as to make the Hamiltonian matrix well structured, viz., real for all binary double groups in the case of an even number of electrons and
quaternion (for $C_1^\ast$ and $C_i^\ast$; cf. \eqref{HCase1}), complex (for $C_2^\ast$, $C_s^\ast$ and $C_{2h}^\ast$; cf. \eqref{HCase2}) or
real (for $C_{2v}^\ast$, $D_2^\ast$ and $D_{2h}^\ast$; cf. \eqref{HCase3})
in the case of an odd number of electrons. Compared with no use of any symmetry, the computational cost can be reduced by a
factor of $|G^\ast|=2|G|$ (NB: due to uneven distributions of the symmetrized functions among the irreps,
 the actual reduction of the computational cost may be somewhat less than this idea factor).
It has been shown\cite{SpinUGA1} that, for an odd number of electrons, the Hamiltonian matrix can also be made real-valued
for $C_1^\ast$, $C_i^\ast$, $C_2^\ast$, $C_s^\ast$, and $C_{2h}^\ast$
by adding in a non-interacting electron. However, the CI space is then doubled (because of the spin-up and spin-down couplings of the extra electron, cf.  Eq. \eqref{twoSval})
and hence has no particular advantage over the direct use of complex algebra.

\section{SOiCI and iCISO}\label{SecSOiCI}
Having introduced the two-step sf-X2C+so-DKH1 Hamiltonian and symmetrized many-electron basis functions,
it is necessary to outline the iCIPT2 approach\cite{iCIPT2,iCIPT2New} for accurate
descriptions of strongly correlated electrons\cite{BlindTest,SDSRev}. Like other selected CI,
iCIPT2 proceeds in two steps, selection of important configurations for static correlation and perturbative treatment of
the first-order interacting space $Q$ for dynamical correlation.
%However, iCIPT2 is unique in the use of CSFs as the many-electron basis
%and the tabulated unitary group approach (TUGA)\cite{iCIPT2} for fast evaluation and reuse of the basic coupling coefficients between selected CSF pairs
%of many open-shell orbitals, which allow an easy extension of iCIPT2 to include SOC.
As stated before, only slight modifications of the TUCA code\cite{iCIPT2} are necessary to handle SOC.
Since different components of a CSF $|J\mathbf{d}\rangle$ are generally mixed by spin-orbit interaction, it is natural to
include all the components $\{J\mathbf{d}\mu\rangle\}$ of a double group and time reversal adapted CSF
once one component is selected. Moreover, care should be take of the invariance of a degenerate manifold
$\{E_k^{(0)}; |\Psi_{k,l}^{(0)}\rangle=\sum_{J\mathbf{d}\mu}|J\mathbf{d}\mu\rangle C^{(0)}_{|J\mathbf{d}\mu\rangle,k,l}\}_{l=1}^{\mathcal{N}_k}$
 of state $k$ upon a unitary transformation within the manifold. Keeping these restrictions to
the reference space $P_0$, the selection criteria can be set up as follows.

A configuration $|J\rangle\in P_0$ of spin $S$ can interact with a configuration $|I\rangle\in Q=1-P_0$ of spin $S'$ ($=S, S\pm 1$),
with the interaction matrix elements being $\langle I\mathbf{d}'\mu'|H|J\mathbf{d}\mu\rangle \equiv H^{I\mathbf{d}'J\mathbf{d}}_{\mu'\mu}$.
Note that the diagonal elements $H^{I\mathbf{d}'I\mathbf{d}'}_{\mu'\mu'}$ reduce to $(H_{sf})^{I\mathbf{d}'I\mathbf{d}'}_{\mu'\mu'}\equiv\mathbb{H}^{I\mathbf{d}'I\mathbf{d}'}$
(independent of $\mu'$), because $(H_{so})^{I\mathbf{d}'I\mathbf{d}'}_{\mu'\mu'}\equiv 0$ due to the antisymmetry of $H_{so}$.
To simplify the notation, we further define the following quantities
\begin{eqnarray}
    A^{I\mathbf{d'}J\mathbf{d}}_{sd,k}
    & = & \sqrt{ \frac{1}{\mathcal{N}_k}
    \sum_{l=1}^{\mathcal{N}_k}\sum_{\mu'=1}^{n_{\gamma'}^{\mathbf{d}'}}\left|\sum_{\mu=1}^{n_{\gamma}^{\mathbf{d}}}\langle I\mathbf{d}'\mu' |H_{sd}| J\mathbf{d}\mu\rangle C_{|J\mathbf{d}\mu\rangle,k,l}^{(0)}\right|^2 }\nonumber\\
    & = & \sqrt{\frac{1}{\mathcal{N}_k}\operatorname{Tr}(\mathbf{C}_k^{\mathbf{d}\dagger} (\mathbf{H}_{sd}^{I\mathbf{d}'J\mathbf{d}})^\dagger\mathbf{H}_{sd}^{I\mathbf{d}'J\mathbf{d}}\mathbf{C}_k^{\mathbf{d}}}), \quad (\mathbf{C}^{\mathbf{d}}_k)_{\mu,l} = C_{|J\mathbf{d}\mu\rangle,k,l}^{(0)},\label{A_ddsd}\\
	A^{I\mathbf{d'}J\mathbf{d}}_{sf,k} &=&  \sqrt{ \frac{1}{\mathcal{N}_k}
    \sum_{l=1}^{\mathcal{N}_k}\sum_{\mu'=1}^{n_{\gamma}^{\mathbf{d}}}\left|\sum_{\mu=1}^{n_{\gamma}^{\mathbf{d}}}\langle I\mathbf{d}'\mu' |H_{sf}| J\mathbf{d}\mu\rangle C_{|J\mathbf{d}\mu\rangle,k,l}^{(0)}\right|^2 }\nonumber\\
&=&\sqrt{\frac{1}{\mathcal{N}_k}\operatorname{Tr}(\mathbf{C}_k^{\mathbf{d}\dagger} (\mathbf{H}_{sf}^{I\mathbf{d}'J\mathbf{d}})^\dagger\mathbf{H}^{I\mathbf{d}'J\mathbf{d}}_{sf}\mathbf{C}^{\mathbf{d}}_k}),\quad \mathbf{H}^{I\mathbf{d}'J\mathbf{d}}_{sf}=\mathbb{H}^{I\mathbf{d}'J\mathbf{d}}_{sf}\mathbf{I}_{n_{\gamma}^{\mathbf{d}}},\label{A_ddsf}
\end{eqnarray}
which are obviously invariant with respect to unitary transformations of the degenerate manifold of state $k$ due to the averaging.
%Given the weak interplay in between correlation and SOC, one can
%separate the quantity in Eq. \eqref{A_dd} into a spin-free and a spin-dependent term
%\begin{subequations}
%	\begin{equation}
%	A^{I\mathbf{d'}J\mathbf{d}}_{sf,k} = \sqrt{\frac{1}{\mathcal{N}_k}\operatorname{Tr}(\mathbf{C}_k^{\mathbf{d}\dagger} (\mathbf{H}_{sf}^{I\mathbf{d}'J\mathbf{d}})^\dagger\mathbf{H}^{I\mathbf{d}'J\mathbf{d}}_{sf}\mathbf{C}^{\mathbf{d}}_k}),\\
%	\end{equation}
%	\begin{equation}
%	A^{I\mathbf{d'}J\mathbf{d}}_{sd,k} = \sqrt{\frac{1}{\mathcal{N}_k}\operatorname{Tr}(\mathbf{C}_k^{\mathbf{d}\dagger} (\mathbf{H}_{sd}^{I\mathbf{d}'J\mathbf{d}})^\dagger\mathbf{H}^{I\mathbf{d}'J\mathbf{d}}_{sd}\mathbf{C}^{\mathbf{d}}_k}),
%	\end{equation}
%\end{subequations}
%thereby neglecting the cross terms between $H_{sf}$ and $H_{so}$.
Following the previous iCI criteria\cite{iCIPT2New}, we then have
\begin{enumerate}[(A)]
	\item\label{Case1} If $|I\rangle$ is identical with or singly excited from $|J\rangle$, then
	\begin{equation}
	\max_{\mathbf{d}}\max_k A^{I\mathbf{d'}J\mathbf{d}}_{sf,k} \geq C_{\mathrm{min}}
	\text{ and }
	\max_{\mathbf{d}}\max_k\frac{A^{I\mathbf{d'}J\mathbf{d}}_{sf,k}}{E_k^{(0)}-\mathbb{H}^{I\mathbf{d}'I\mathbf{d}'}} \geq C_{\mathrm{min}}, \label{iCI_Rank_1_1}		
	\end{equation}
	or
	\begin{equation}
	\max_{\mathbf{d}}\max_k A^{I\mathbf{d'}J\mathbf{d}}_{sd,k} \geq C_{\mathrm{min}}
	\text{ and }
	\max_{\mathbf{d}} \max_k\frac{A^{I\mathbf{d'}J\mathbf{d}}_{sd,k}}{E_k^{(0)}-\mathbb{H}^{I\mathbf{d}'I\mathbf{d}'}} \geq C_{\mathrm{min}}. \label{iCI_Rank_1_2}		
	\end{equation}	
	\item\label{Case2} If $|I\rangle$ is doubly excited from $|J\rangle$, only $H_{sf}$ is involved, then
	\begin{eqnarray}
	&&	\max_{\mathbf{d}} \max_k
	\tilde{H}^{IJ}\sqrt{\frac{1}{\mathcal{N}_k}\operatorname{Tr}(\mathbf{C}_k^{\mathbf{d}\dagger}\mathbf{C}_k^\mathbf{d})}
	\geq C_{\mathrm{min}}\text{ and }\max_{\mathbf{d}}\max_k A^{I\mathbf{d'}J\mathbf{d}}_{sf,k} \geq C_{\mathrm{min}}\nonumber\\
	&&\text{ and } \max_{\mathbf{d}}\max_k \frac{A_{sf,k}^{I\mathbf{d'}J\mathbf{d}}}{E_k^{(0)}-\mathbb{H}^{I\mathbf{d}'I\mathbf{d}'}} \geq C_{\mathrm{min}}. \label{iCI_Rank_2}
	\end{eqnarray}
\end{enumerate}

In detail, for case (\ref{Case1}), loop over all symmetry adapted CSFs $|I\mathbf{d}'\rangle$
associated with configuration $|I\rangle\in Q$ and evaluate $A^{\mathbf{d}'\mathbf{d}}_{sf,k}$ and $A^{\mathbf{d}'\mathbf{d}}_{sd,k}$ for all CSFs $ |J\mathbf{d}\rangle\in P_0$.
If $\max_{\mathbf{d}}\max_k A^{\mathbf{d'}\mathbf{d}}_{sf,k}$ or $\max_{\mathbf{d}}\max_k A^{\mathbf{d'}\mathbf{d}}_{sd,k}$ is larger than $C_{\mathrm{min}}$ then evaluate the diagonal matrix elements $\mathbb{H}^{I\mathbf{d}'I\mathbf{d}'}$;
otherwise discard $|I\mathbf{d}'\rangle$. If $\max_{\mathbf{d}}\max_k\frac{A_k^{I\mathbf{d'}J\mathbf{d}}}{E_k^{(0)}-\mathbb{H}^{I\mathbf{d}'I\mathbf{d}'}}$ is larger than $C_{\mathrm{min}}$ then $|I\mathbf{d}'\rangle$ is selected.

As for case (\ref{Case2}), only those doubly excited configurations $|I\rangle$ with the estimated, CSF-independent two-body integrals
$\tilde{H}^{IJ}$ (see Ref. \citenum{iCIPT2}) larger than $C_{\mathrm{min}}/\max_k\sqrt{\frac{1}{\mathcal{N}_k}\operatorname{Tr}(\mathbf{C}_k^{\mathbf{d}\dagger}\mathbf{C}_k^\mathbf{d})}$ need to be generated
(i.e., those unimportant ones are never touched, just like determinant-based heat-bath CI\cite{HBCI2016}). For such $\{|I\rangle\}$, the remaining step is the same as case (\ref{Case1}).
%It deserves to be pointed out that in this case, since the two-body operator is spin-free, $S'=S$ and
%$\mathbf{H}^{I\mathbf{d'}J\mathbf{d}} = \langle I\mathbf{d}'|H_{sf}|J\mathbf{d}\rangle \mathbf{I}$ where $\mathbf{I}$ is the identity matrix of order $2S+1$.

The above ranking procedure expands the reference $P_0$ to $P_1$, which is diagonalized by the iterative vector interaction (iVI) approach\cite{iVI,iVI-TDDFT}.
Those CSFs of coefficients smaller in absolute value than $C_{\mathrm{min}}$ are then pruned away, leading to $P$.
The procedure is iterated until $P$ and $P_0$ are sufficiently similar in compositions. It has been shown\cite{iCIPT2New} that
such combined integral- and coefficient-driven, ranking-pruning selection scheme
is highly efficient in building up iteratively a compact variational space $P$.

What has been described so far is a procedure that selects important CSFs for both static correlation and SOC. Given the weak interplay in between,
one can simply invoke conditions \eqref{iCI_Rank_1_1} and \eqref{iCI_Rank_2} to select iteratively important CSFs of spins $S-1$, $S$ and $S-1$ for static correlation alone and then
invoke condition \eqref{iCI_Rank_1_2} to select non-iteratively additional singly excited CSFs important for SOC.
Although not documented here, it has been confirmed numerically that such combined and separate selections lead to
virtually identical results for systems considered here. Therefore, the latter will be used throughout.

Upon termination of the selection, the residual dynamic correlation is estimated by using the state-specific Epstein-Nesbet type of
second-order perturbation theory (ENPT2):
\begin{eqnarray}
E_{c,k,l}^{(2)} &=& \sum_{|I\mathbf{d}'\mu'\rangle\in Q}\frac{|\langle I\mathbf{d}'\mu'|H|\Psi^{(0)}_{k,l}\rangle|^2}{E^{(0)}_k-\mathbb{H}^{I'\mathbf{d}'I'\mathbf{d}'}}\label{ENPT2Defa}\\
&=&\sum_{|I\mathbf{d}'\mu'\rangle\in Q} \frac{\left|\sum_{|J\mathbf{d}\mu\rangle\in P}H^{I\mathbf{d}'J\mathbf{d}}_{\mu'\mu}C^{(0)}_{|J\mathbf{d}\mu\rangle,k,l}\right|^2}{E_k^{(0)}-\mathbb{H}^{I'\mathbf{d}'I'\mathbf{d}'}}.\label{ENPT2Defb}
\end{eqnarray}
which can be reexpressed as\cite{ASCI2018PT2}
\begin{align}
E_{c,k,l}^{(2)} &=\tilde{E}_{c,k,l}^{(2)}-\bar{E}_{c,k,l}^{(2)},\label{ENPT2diff}\\
\tilde{E}_{c,k,l}^{(2)}&= \sum_{|I\mathbf{d}'\mu'\rangle\in P\cup Q} \frac{\left|\sum_{|J\mathbf{d}\mu\rangle\in P, |J\mathbf{d}\mu\rangle\ne |I\mathbf{d}'\mu'\rangle}H^{I\mathbf{d}'J\mathbf{d}}_{\mu'\mu}C_{|J\mathbf{d}\mu\rangle,k,l}^{(0)}\right|^2}{E_k^{(0)}-\mathbb{H}^{I\mathbf{d}'I\mathbf{d}'}},\label{ENPT2Qt}\\
\bar{E}_{c,k,l}^{(2)}&=\sum_{|J\mathbf{d}\mu\rangle\in P} \frac{\left|\langle J\mathbf{d}\mu|H|\Psi_{k,l}^{(0)}\rangle-\mathbb{H}^{J\mathbf{d}J\mathbf{d}}C_{|J\mathbf{d}\mu\rangle,k,l}^{(0)} \right|^2}{E_k^{(0)}-\mathbb{H}^{J\mathbf{d}J\mathbf{d}}}\label{ENPT2Pa}\\
&=\sum_{|J\mathbf{d}\mu\rangle\in P} |C^{(0)}_{|J\mathbf{d}\mu\rangle,k,l}|^2 (E_k^{(0)}-\mathbb{H}^{J\mathbf{d}J\mathbf{d}}),\label{ENPT2P}
\end{align}
where
%$|\mu\rangle$ is short for $|I\mathbf{d}\rangle$ and
the use of the relation $\langle J\mathbf{d}\mu|H|\Psi_{k,l}^{(0)}\rangle=C_{|J\mathbf{d}\mu\rangle,k,l}^{(0)} E_k^{(0)}$ has been made when going from Eqs. \eqref{ENPT2Pa} to \eqref{ENPT2P}.
This reformulation avoids the expensive double check of the excited CSFs belonging to Q or P.
The use of pre-ordered arrays\cite{iCIPT2New} for Timsorting\cite{Timsort,TimsortCode} the residues\cite{Residue,ASCI2020}
involved in the constraint-based\cite{ASCI2018PT2} ENPT2 allows
allows a massive parallel implementation.

\begin{figure}
\centering
	\includegraphics[width=1.0\textwidth]{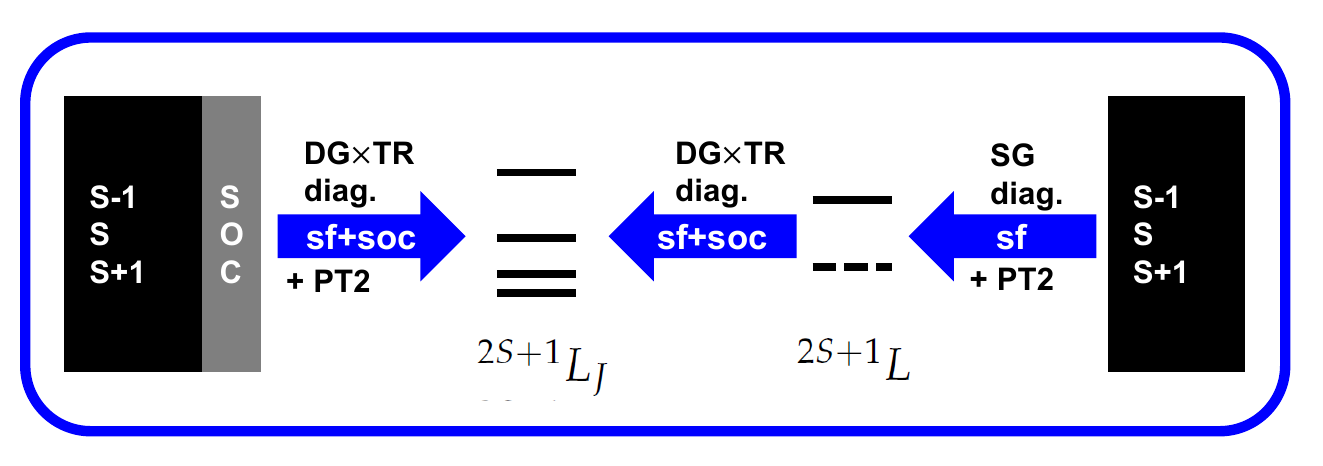}
	\caption{Flowchart of SOiCI (left to right) and iCISO (right to left).}
	\label{FigSOiCI}
\end{figure}

The above formulation defines SOiCI, which treats electron correlation and SOC on an equal footing
and hence belongs to the one-step family of methods outlined in the Introduction. It can be simplified in two ways:
(1) only those CSFs with coefficients larger in absolute value than $C_{\mathrm{min}}^{\mathrm{SO}}$ (e.g.,
$10^{-4}$, which is larger $C_{\mathrm{min}}$) are employed as references to select additionally singly excited CSFs for SOC with condition \eqref{iCI_Rank_1_2}.
Although the ENPT2 correlation corrections have to be evaluated twice (one for sf-X2C-iCIPT2 with $C_{\mathrm{min}}$ and the other for SOiCI
with $C_{\mathrm{min}}^{\mathrm{SO}}$),
there is still gain in efficiency when  $C_{\mathrm{min}}$ is much smaller than $C_{\mathrm{min}}^{\mathrm{SO}}$.
Moreover, if wanted, only the sf-X2C-iCIPT2 energies need to be extrapolated with different $C_{\mathrm{min}}$.
This Ansatz may be termed intermediate SOiCI (iSOiCI).
(2) A more dramatic simplification follows directly the idea of state interaction, that is, an effective spin-orbit Hamiltonian matrix in
the basis of a small number of correlated scalar states is constructed
and diagonalized. More specifically, the diagonal elements herein are the sf-X2C-iCIPT2 energies, whereas the off-diagonal elements
are the spin-orbit interactions between the zeroth-order wave functions within the P space selected
by the spin-free counterpart\cite{iCIPT2New} of conditions \eqref{iCI_Rank_1_1} and \eqref{iCI_Rank_2}. Note that
iCISO cannot be viewed simply as a contracted version of SOiCI since the spin-dependent selection of configurations with condition \eqref{iCI_Rank_1_2}
is not invoked therein. An illustration of SOiCI and iCISO is plotted in Fig. \ref{FigSOiCI}.
%in the latter it is assumed that
%SOC and electron correlation are additive, such that just a few number of wave functions of nonrelativistic symmetry is enough to expand the relativistic wave function.
%As a matter of fact, there exists an intermediate approach between the two extremes\cite{Teichteil2000},
%where the effective Hamiltonian matrix is defined on a set of configuration state functions (CSF) or determinants (SD), the number of which
%is much smaller than those in the one-step approaches but much larger than the (contracted) scalar states in the two-step approaches.
%The three would agree with each other completely if all the CSFs/SDs or equivalently linear combinations of them from the same active space are taken into account.

\section{Results and discussion}\label{Results}
\subsection{Generality}
It is well known that the heavy $p$-block elements in the Periodic Table are most challenging to the treatment of SOC,
because the $np_{1/2}$ spinors are weakly singular in the vicinity of the nuclear position and have very different spatial extensions from $np_{3/2}$.
While these features can readily be resolved if SOC is treated variationally already at the mean-field level, they do impose a serious issue when SOC is postponed
to the correlation step, for the chosen (contracted) scalar $np$ orbitals are close to $np_{3/2}$ but are very different from $np_{1/2}$ in spatial extensions.
This means essentially that single excitations to the `right' virtual orbitals (e.g., those with nodes close to the maximum of $np$ and extrema on the two sides)
ought to be captured, so as to bring in biased polarization of $np$ towards $np_{1/2}$ (i.e., spin-dependent orbital relaxation).
%This means that the deficiency of using scalar $p$ orbitals to describe the $np_{1/2}$ spinors must be accounted for by single excitations to
%suitable virtual orbitals
The question is how to identify such `right' virtual orbitals in an automatic manner. In principle, one can perform
a selection of configuration with some criterion involving the SOC matrix elements, e.g., condition \eqref{iCI_Rank_1_2} adopted here.
However, such selection may be dangerous. For instance, when working with an uncontracted basis with steep functions,
very likely some virtual orbitals
that are very high in energy but are very local (HELO) in space will be picked up (because of appreciable
spin-orbit integrals), but they correspond to `anti-bonding' orbitals of innermost core orbitals
and therefore have nothing to do with the spin-orbit splitting (SOS) of a valence $np$ orbital. As a matter of fact,
including such HELOs usually worsens the result if the innermost core orbitals are not
included in the treatment of electron correlation (which is usually the case).
To avoid such situation, we borrow the idea of virtual space decomposition (VSD)\cite{SDSRev} designed originally to reduce
the correlation space. Specifically, a small set of orthonormal AOs $\{\tilde{\chi}_{\mu}\}_{\mu=1}^{N^S}$, derived by symmetric orthonormalization of, e.g., a SRECP (scalar relativistic effective core potential) double-zeta basis,
can be used to map out a reduced set of virtual orbitals $\{\tilde{\psi}_a^L\}$ from the all-electron virtual orbitals $\{\psi_b\}_{b=1}^{N_{vir}^L}$, by a singular value decomposition of the overlap
 \begin{equation}
 \langle\tilde{\chi}_{\mu}|\psi_a\rangle=(\mathbf{L}\boldsymbol{\lambda}\mathbf{R}^\dag)_{\mu a},\label{VSDa}
 \end{equation}
followed by the rotation of the virtual orbitals
\begin{equation}
 |\tilde{\psi}_a^L\rangle=\sum_{b=1}^{N_{vir}^L} |\psi_b\rangle R_{ba}.\label{VSDb}
 \end{equation}
Those $\{|\tilde{\psi}_a^L\rangle\}$ with $\tilde{N}^L_{vir}=N^S+N_{core}-N_{occ}$ largest singular values $\lambda_a$ are then taken as the effective virtual orbitals.
Here, $N_{occ}$ is the total number of doubly occupied and active orbitals in the all-electron calculation.
In this way, both the $N_{core}$ innermost core orbitals (corresponding to those replaced by SRECP) and their `anti-bonding' virtual orbitals are
removed, leaving a set of all-electron orbitals that are suitable for both correlation and SOC.
Another issue that deserves attention lies in that SOC and correlation require different configuration spaces to describe, for they have very different physical origins
(one-body spin-orbit vs two-body Coulomb interaction).
For instance, it is often the case that a good result for the SOS
of a valence $np$ configuration can be obtained by correlating just a small number of electrons in a small number of valence orbitals, but the
result may be deteriorated significantly when the number of correlated electrons or orbitals is increased individually\cite{SOCsingle-1,CI-SOC1998,Teichteil2000}.
In short, the number of electrons (or occupied orbitals) as well as the number and character of virtual orbitals must be chosen carefully to achieve a balanced description of SOC and correlation.

To confirm the above points on one hand and test the proposed SOiCI and iCISO methods on the other,
we take both atomic (Sec. \ref{Atoms}) and diatomic (Sec. \ref{Diatomics}) systems as examples.
Spin-free X2C state-averaged (SA) CASSCF calculations with the (contracted) relativistic ANO-RCC basis sets\cite{ANO2} were performed to generate the MOs,
under the $D_{2h}$ and $C_{2v}$ symmetries for the atoms and molecules, respectively.
For CAS(12e,12o) and smaller active spaces, no selection was performed, whereas for active spaces larger than  CAS(12e,12o),
the iCISCF approach\cite{iCISCF} (which employs selected iCI\cite{iCI} as the active space solver) was adopted to obtain near-exact CASSCF solutions.
When necessary, the SOiCI energies were extrapolated linearly with five $C_{\mathrm{min}}$ values $\{10.0, 7.0, 5.0, 3.0, 1.5\}\times10^{-5}$.
The standard deviation for the extrapolated spin-orbit splitting is estimated according to $\sigma=\sqrt{\sigma_0^2+\sigma_1^2}$, with $\sigma_0$ and $\sigma_1$
being the standard deviations for the extrapolated energies of the ground and excited spinor states, respectively. It deserves to be pointed out that dynamic
correlation beyond the chosen active spaces was not accounted for, such that SOiCI/iCISO should in principle be denoted\cite{iCISCF} as SOiCI(2)/iCISO(2) to emphasize
the `inner space' second-order dynamic correlation. However, to avoid over notation, such `convention' is not adopted here.
All calculations were performed with the BDF program package\cite{BDF1,BDF2,BDF3,BDFrev2020}.

%The relative weights of determinants in the CI expansion are determined
%solely by the electron repulsion operator, whereas the
%spin-orbit operator shows completely different properties. For instance, single excitations are relatively
%unimportant for $1/r_{ij}$, whereas they are significant
%contributors to spin-orbit matrix elements. Leaving
%them out would result in underestimation of the spin-orbit coupling. It is also to be expected that the amount
%of correlation energy will differ between, for instance, a
%$p_{1/2}$ and a $p_{3/2}$ shell when the spin-orbit coupling is large,
%requiring a simultaneous treatment of the two effects.
\subsection{SOS of $p$-block Atoms}\label{Atoms}
The halogen atoms (F to I) as well as the Pb and Bi atoms have very simple ground states resulting from the open-shell $np$ orbital
and are therefore ideal systems for examining the previous general arguments.
The simplest active space for the halogen atoms is obviously CAS(5e,3o) corresponding to $np^5$ ($n=2$ to 5 for F to I, respectively).
As can be seen from Table \ref{Halogen_Results},
the CAS(5e,3o)-SOiCI results for the SOS of the inversion-odd $^2P^{\circ}$ state
are already in semi-quantitative agreement with (ca. $7\%$ lower than) the experimental data\cite{NIST1}, even for I.
As expected, further including $ns^2$ on top of CAS(5e,3o), i.e., CAS(7e,4o), causes no changes.
The results are also not much improved by further
including the $(n+1)s(n+1)p$ virtual orbitals into the active space (i.e., CAS(7e,8o)) and were even worsened by further including all the rest virtual orbitals,
except for I, for which the result is indeed improved slightly.
This shows clearly that keeping the number of correlated electrons fixed to a very limited value while
increasing the number of virtual orbitals is generally not appropriate for SOS. The next try is then to augment CAS(7e,8o) with
the $(n-1)$-shell electrons and orbitals as well as the mirror $(n+1)d$ virtual orbitals, leading to
CAS(13e,16o) for Cl and CAS(17e,18o) for both Br and I.
As can be seen from Table \ref{Halogen_Results}, the results are much better than those by the CAS with 7 electrons in all virtual orbitals,
reflecting the importance of core-valence polarization effects on SOS.
The results are further improved by including all the rest virtual orbitals, especially for Br and I.
However, including further the $(n-1)p^6$ shell tends to reduce the already underestimated SOS of I.
%which should be ascribed to the deficiency in the treatment of correlation.
%\textcolor[rgb]{1.00,0.00,0.00}{
To see if this is due to the use of a contracted basis, we further performed calculations with the $p$ functions in ANO-RCC decontracted (denoted as p-ANO-RCC).
%At this stage, one should check to what extent the contraction of a basis set
%affects the SOS. Since the SOS here stems primarily from the $p$ orbitals, only the $p$-functions of the ANO-RCC basis are decontracted here (denoted as p-ANO-RCC).
The SOS of I by correlating 17 (23) electrons in all virtual orbitals becomes 8035 (7827) cm$^{-1}$, which is
too large. As noted before, this is due to the presence of HELOs, which are not compensated for by
correlating the innermost core orbitals. Removing such HELOs by the VSD (cf. Eqs. \eqref{VSDa} and \eqref{VSDb})
leads to 7336 (7200) cm$^{-1}$, which is smaller than 7416 (7236) cm$^{-1}$ with the original ANO-RCC basis. This is because the
steepest $p$-functions in p-ANO-RCC have somewhat less contributions to the valence $p$ orbitals than those in ANO-RCC.
Therefore, to keep a balance between correlation and SOC, the $(n-1)d^{10}ns^2np^6$ electrons of I (and Br) should be correlated in a generally contracted basis.

\begin{figure}
	\centering
	\includegraphics[width=1.0\textwidth]{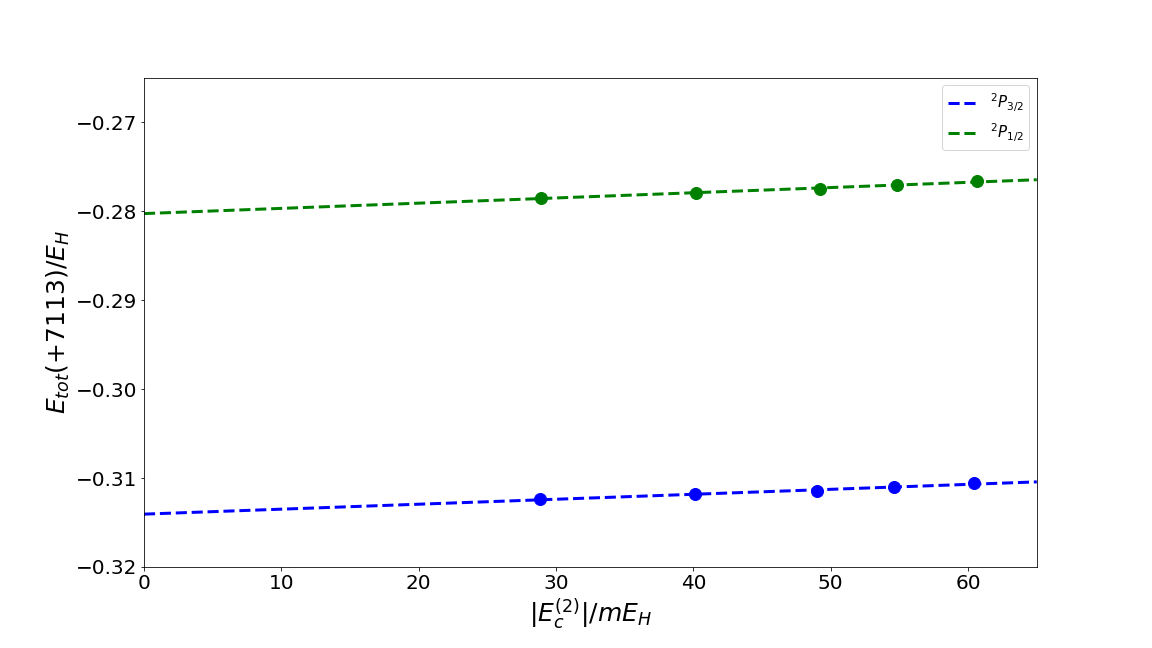}
	\caption{Linear extrapolations of the SOiCI energies of $^2\mathrm{P}_{1/2}$ and $^2\mathrm{P}_{3/2}$ of iodine
		versus the absolute second-order correlation energies $|E_c^{(2)}|$ (17 electrons in all virtual orbitals).}
	\label{extrapolation}
\end{figure}

For more details, the SOiCI results calculated with different values for $C_{\mathrm{min}}$ are shown in Table \ref{Detail_Data_Cl_Br_I_SOiCI}.
It is first seen that the SOS is very insensitive to $C_{\mathrm{min}}$, thereby substantiating the iSOiCI approach (which is only briefly
mentioned in the end of Sec. \ref{SecSOiCI} and will be discussed in more detail elsewhere). Moreover,
the nearly perfect linear relations (cf. Fig. \ref{extrapolation} for I)
 between the SOiCI energies of $^2P_{3/2}$ and $^2P_{1/2}$
and the absolute second-order correlation energies $|E_c^{(2)}|$ allows accurate extrapolations to the zero $C_{\mathrm{min}}$ limit.

As for iCISO, the very first choice is to project the so-DKH1 Hamiltonian onto the degenerate manifold of the lowest $^2P^{\circ}$ state.
Not surprisingly, as a very cheap method, iCISO performs very well for the lightest atoms (F and Cl) but fails for the heaviest ones, with Br
being the borderline. This is of course due to the lack of spin-dependent orbital relaxation, just like other two-step approaches.
Nonetheless, iCISO may perform well for heavy elements when SOC is quenched to a large extent by chemical bondings\cite{sf-X2C-EOM-SOC2017}.
Another point that deserves to be addressed is the sensitivity of iCISO on the number of scalar states.
At first glance, the scalar states arising from $ 5s^1 5p^6$ ($^2S$) and/or $5s^2 5p^4 6s$ ($^4P$, $^2P$, $^2D$, and $^2S$) should first be added.
However, such states are inversion even and hence do not interact with
$^2P^{\circ}$ via SOC. On the other hand, both the $^4P^{\circ}$ and $^4D^{\circ}$ states of $ 5s^2 5p^4 6p$ are energetically too high
to have discernible couplings with $^2P^{\circ}$.

The SOS of \ce{Pb} is more difficult to describe. First of all, the $^3P$, $^1D$, and $^1S$ states of
$5d^{10}6s^26p^2$ are strongly coupled by SOC and should hence all be included during the selection procedure.
Secondly, as indicated by the significant differences between the SOiCI(20e,21o) results with orbitals optimzed by
SA-iCISCF(14e,18o) (see Table \ref{Pb_14e_orb}) and SA-iCISCF(20e,21o) (see Table \ref{Pb_20e_orb}),
the $5p^6$ shell should also be included in the orbital optimization, i.e., iCISCF(20e,21o) averaged (equally) over the $^3P$, $^1D$, and $^1S$ states of $5p^65d^{10}6s^26p^27s7p6d$.
Including all virtual orbitals in the SOiCI calculations then
yields the best estimate of the SOS of Pb, with the maximal error being ca. $2.5\%$.
Similarly, the $^4S$, $^2P$, and $^2D$ states of $5p^65d^{10}6s^26p^3$ should all be included in the SOiCI calculations of Bi.
As can be seen from Table \ref{Bi_21e_orb}, the SA-iCISCF(21e,21o) orbital-based SOiCI SOS of Bi, with all virtual orbitals included,
are in very good agreement with experiments. 

Finally, a brief comparison should be made between SOiCI and other related approaches\cite{HBCISOC2017,sf-X2C-EOM-SOC2017}.
As can be seen from Table \ref{Atom_Extrapolation} for the halogen atoms, only
the SOSHCI (spin-orbit stochastic heat-bath CI) value\cite{HBCISOC2017} for the SOS of Br is an outlier to the good agreement in between.
Since SOSHCI employs the same two-step sf-X2C+so-DKH1 Hamiltonian (see Sec. \ref{SecsfX2C}), their calculations on Br should be repeated.
As for Pb, the SOiCI results are also in good agreement with the previous theoretical values\cite{ccpVDZPP1}.

\begin{table}
	\caption{
		Spin-orbit splitting (in cm$^{-1}$) of halogen atoms calculated by SOiCI and iCISO with the ANO-RCC basis sets and various active spaces.
In parentheses are standard deviations of linear extrapolations (cf. Table \ref{Detail_Data_Cl_Br_I_SOiCI})
	}
	\centering
	\begin{threeparttable}
		\centering
		\begin{tabular}{c|c|ccllll|c}\toprule
			%			\multicolumn{4}{c}{\ce{F}} \\
			\multirow{1}{*}{atom}
			& \multirow{1}{*}{method}
			& \multicolumn{6}{c|}{active space}
			& Expt.\tnote{h}\\\toprule
			%			& (7e,4o) & (7e,8o) & (7e,87o) \\ \toprule
			\multirow{3}{*}{\ce{F}}
			&
%			&(7e,4o)\tnote{a}
			&(5e,3o)\tnote{a}
			&(7e,8o)\tnote{b}
			&(7e,87o)\tnote{c}
			&   &   &   &\\\cline{2-8}
			& SOiCI
			& 405
			& 401
			& 398(1)
			
			&&&
			&\multirow{2}{*}{404}\\
			& iCISO
			& 405
			& 401
			& 398
			&   &   &   &\\\midrule
			\multirow{3}{*}{\ce{Cl}}
			&
%			&(7e,4o)\tnote{a}
            &(5e,3o)\tnote{a}
			&(7e,8o)\tnote{b}
			&(7e,95o)\tnote{c}
			& (13e,16o)\tnote{d}
%			\tnote{d}
%			& (17e,100o)
			& (13e,98o)\tnote{e}
%			\tnote{d}
			&
			&\\\cline{2-8}
			& SOiCI
			& 829
			& 837
			& 807(3)
			& 888
%			& 871(8)
			& 871(3)
			&
			&\multirow{2}{*}{882}\\
			& iCISO
			& 829
			& 835
			& 805
			& 887
%			& 868
            & 869
			&
			&\\\midrule
			\multirow{3}{*}{\ce{Br}}
			&
%			&(7e,4o)\tnote{a}
            &(5e,3o)\tnote{a}
			&(7e,8o)\tnote{b}
			&(7e,95o)\tnote{c}
			&(17e,18o)\tnote{f}
			&(17e,100o)\tnote{g}
			&(23e,103o)\tnote{g}
			&\\\cline{2-8}
			& SOiCI
			& 3429
			& 3479
			& 3378(1)
			& 3792
			& 3745(58)
			& 3642(85)
			&\multirow{2}{*}{3685}\\
			& iCISO
			& 3429
			& 3434
			& 3298
			& 3708
			& 3643
			& 3582
			&\\\midrule
			\multirow{3}{*}{\ce{I}}
			&
%			&(7e,4o)\tnote{a}
			&(5e,3o)\tnote{a}
			&(7e,8o)\tnote{b}
			&(7e,116o)\tnote{c}
			&(17e,18o)\tnote{f}
			&(17e,121o)\tnote{g}
			&(23e,124o)\tnote{g}
			&\\\cline{2-8}
			& SOiCI
			& 7024
			& 7138
			& 7144(1)
			& 7135
			& 7416(38)
			& 7236(36)
			&\multirow{2}{*}{7603}\\
			& iCISO
			& 7024
			& 6938
			& 6670
			& 7137
			& 7047
			& 7143
			&\\\midrule
		\end{tabular}
		\begin{tablenotes}
			\item[a] CASSCF(5e,3o) orbitals (identical with CASSCF(7e,4o)).
			\item[b] CASSCF(7e,8o) orbitals.
			\item[c] Including all virtual orbitals from CASSCF(7e,8o).
			\item[d] iCISCF(13e,16o) orbitals.
			\item[e] Including all virtual orbitals from iCISCF(13e,16o).
			\item[f] iCISCF(17e,18o) orbitals.
			\item[g] Including all virtual orbitals from iCISCF(17e,18o).
			\item[h] Experiments\cite{NIST1}.
		\end{tablenotes}
	\end{threeparttable}
	\label{Halogen_Results}
\end{table}

\begin{table}
	\small
	\centering
	\caption{Spin-orbit splitting (in cm$^{-1}$) of halogen atoms calculated by SOiCI with the largest active spaces in Table \ref{Halogen_Results} and various $C_{\mathrm{min}}$ values
	}		
	\begin{threeparttable}
		\begin{tabular}{c|c|cccc|c}\toprule
			\multirow{2}{*}{atom}
			& \multirow{2}{*}{$C_{\mathrm{min}}$}
			& \multicolumn{2}{c}{$^{2}P_{3/2}$}
			& \multicolumn{2}{c|}{$^{2}P_{1/2}$}
			& \multirow{2}{*}{splitting} \\ \cline{3-6}
			&
			&$E_c^{(2)}/\mathrm{mE_{H}}$&$E_{tot}/\mathrm{E_H}$
			&$E_c^{(2)}/\mathrm{mE_{H}}$&$E_{tot}/\mathrm{E_H}$
			&\\\toprule
			\multirow{7}{*}{F}
			& $1.0\times10^{-4}$ & -4.83  & -99.74368   & -4.83  & -99.74187   & 399  \\
			& $7.0\times10^{-5}$ & -3.95  & -99.74368   & -3.95  & -99.74186   & 399  \\
			& $5.0\times10^{-5}$ & -3.25  & -99.74369   & -3.25  & -99.74188   & 399  \\
			& $3.0\times10^{-5}$ & -2.32  & -99.74371   & -2.32  & -99.74189   & 399  \\
			& $1.5\times10^{-5}$ & -1.37  & -99.74373   & -1.37  & -99.74191   & 398  \\
			& 0.0                &        & -99.74374   &        & -99.74193   & 398(1)  \\\midrule
			\multirow{7}{*}{Cl}
%			& $1.0\times10^{-4}$ & -18.48 & -461.27557  & -18.46 & -461.27160  & 870  \\
%			& $7.0\times10^{-5}$ & -15.19 & -461.27582  & -15.18 & -461.27186  & 871  \\
%			& $5.0\times10^{-5}$ & -12.63 & -461.27601  & -12.62 & -461.27204  & 871  \\
%			& $3.0\times10^{-5}$ & -9.48  & -461.27621  & -9.47  & -461.27224  & 871  \\
%			& $1.5\times10^{-5}$ & -6.46  & -461.27638  & -6.45  & -461.27241  & 871  \\
%			& 0.0                &        & -461.27684  &        & -461.27287  & 871(8)  \\\midrule
			& $1.0\times10^{-4}$ &-16.93  &-461.23207   &-16.93  &-461.22811  &870 \\
			& $7.0\times10^{-5}$ &-13.69  &-461.23266   &-13.69  &-461.22869  &871 \\
			& $5.0\times10^{-5}$ &-11.20  &-461.23310   &-11.21  &-461.22914  &871 \\
			& $3.0\times10^{-5}$ &-8.24   &-461.23362   &-8.24   &-461.22965  &871 \\
			& $1.5\times10^{-5}$ &-5.55   &-461.23410   &-5.55   &-461.23014  &871 \\
			& 0.0                &        &-461.23509   &        &-461.23112  &871(2)  \\\midrule
			\multirow{7}{*}{Br}
            & $1.0\times10^{-4}$ &-21.29  &-2604.75709  &-21.23  &-2604.74014  &3720 \\
            & $7.0\times10^{-5}$ &-17.62  &-2604.75772  &-17.57  &-2604.74075  &3726 \\
            & $5.0\times10^{-5}$ &-15.03  &-2604.75815  &-15.02  &-2604.74117  &3728 \\
            & $3.0\times10^{-5}$ &-11.89  &-2604.75889  &-11.87  &-2604.74190  &3729 \\
            & $1.5\times10^{-5}$ &-8.39   &-2604.75997  & -8.38  &-2604.74298  &3729 \\
            & 0.0                &        &-2604.76163  &        &-2604.74461  &3735(58) \\\midrule
			\multirow{7}{*}{I}
			& $1.0\times10^{-4}$ &-60.44  &-7113.31055  &-60.68  &-7113.27659  &7455 \\
			& $7.0\times10^{-5}$ &-54.62  &-7113.31105  &-54.83  &-7113.27710  &7451 \\
			& $5.0\times10^{-5}$ &-49.03  &-7113.31146  &-49.20  &-7113.27753  &7448 \\
			& $3.0\times10^{-5}$ &-40.08  &-7113.31185  &-40.22  &-7113.27794  &7441 \\
			& $1.5\times10^{-5}$ &-28.83  &-7113.31237  &-28.92  &-7113.27850  &7434 \\
			& 0.0                &        &-7113.31406 &        & -7113.28027 & 7416(38) \\\bottomrule
		\end{tabular}
	\end{threeparttable}
	\label{Detail_Data_Cl_Br_I_SOiCI}
\end{table}

% Pb

\begin{table}[!htp]
	\centering
	\caption{SA-iCISCF(14e,18o) orbital-based SOiCI energies (in cm$^{-1}$)
		for the low-lying states of \ce{Pb} relative to the ground state $^3P_0$.
		In parentheses are standard deviations of linear extrapolations
	}
	\begin{threeparttable}
		\begin{tabular}{l|llll}\toprule
%			\multirow{3}{*}{active space}
%			&\multicolumn{4}{c}{excited spinor states} \\ \cline{2-5}
%			&\multicolumn{2}{c|}{(3/2,1/2)}
%			&\multicolumn{2}{c}{(3/2,3/2)} \\\cline{2-5}
%			& \multicolumn{1}{c}{$J=1$}
%			& \multicolumn{1}{c|}{$J=2$}
%			& \multicolumn{1}{c}{$J=2$}
%			& \multicolumn{1}{c}{$J=0$} \\\toprule
			\multicolumn{1}{c|}{active space}
			& \multicolumn{1}{c}{$^3P_1$}
			& \multicolumn{1}{c}{$^3P_2$}
			& \multicolumn{1}{c}{$^1D_2$}
			& \multicolumn{1}{c}{$^1S_0$} \\ \toprule
			CAS(14e,18o)\tnote{a}
			& 6845(3)
			& 10606(2)
			& 21302(2)
			& 31877(2)\\
			CAS(20e,21o)\tnote{b}
			& 6322(4)
			& 10012(4)
			& 20266(4)
			& 30776(3)\\
			CAS(14e,144o)\tnote{c}
			& 8838(38)
			& 11704(40)
			& 23480(31)
			& 31557(27)\\
			CAS(20e,147o)\tnote{d}
			& 7937(5)
			& 10826(15)
			& 20961(8)
			& 28711(39)\\ \midrule
			\multicolumn{1}{c|}{Expt.\tnote{e}}
			& 7819
			& 10650
			& 21458
			& 29467 \\\bottomrule
		\end{tabular}
		\begin{tablenotes}
			\item[a] 14 electrons in $5d6s6p7s7p6d$ orbitals.
			\item[d] 20 electrons in $5p5d6s6p7s7p7d$ orbitals.
			\item[c] 14 electrons in $5d6s6p$ and all virtual orbitals.
			\item[d] 20 electrons in $5p5d6s6p$ and all virtual orbitals.
			\item[e] Experiments\cite{NIST1}.
		\end{tablenotes}
		\label{Pb_14e_orb}
	\end{threeparttable}
\end{table}

\begin{table}[!htp]
	\centering
	\caption{SA-iCISCF(20e,21o) orbital-based SOiCI energies (in cm$^{-1}$)
		for the low-lying states of \ce{Pb} relative to the ground state $^3P_0$.
		In parentheses are standard deviations of linear extrapolations}
	\begin{threeparttable}
		\begin{tabular}{l|llll}\toprule
%			\multirow{3}{*}{active space}
%			&\multicolumn{4}{c}{excited spinor states} \\ \cline{2-5}
%			&\multicolumn{2}{c|}{(3/2,1/2)}
%			&\multicolumn{2}{c}{(3/2,3/2)} \\\cline{2-5}
%			& \multicolumn{1}{c}{$J=1$}
%			& \multicolumn{1}{c|}{$J=2$}
%			& \multicolumn{1}{c}{$J=2$}
%			& \multicolumn{1}{c}{$J=0$} \\\toprule
			\multicolumn{1}{l|}{active space}
			& \multicolumn{1}{c}{$^3P_1$}
			& \multicolumn{1}{c}{$^3P_2$}
			& \multicolumn{1}{c}{$^1D_2$}
			& \multicolumn{1}{c}{$^1S_0$} \\ \toprule
			CAS(20e,21o)\tnote{a}
			& 6510(4)
			& 10170(4)
			& 20456(4)
			& 30647(4)\\
			CAS(20e,30o)\tnote{b}
			& 6437(4)
			& 10040(4)
			& 20105(4)
			& 29610(3)\\
			CAS(20e,39o)\tnote{c}
			& 6027(6)
			& 9443(5)
			& 19008(4)
			& 27824(3)\\
			CAS(20e,48o)\tnote{d}
			& 6476(3)
			& 9905(3)
			& 19786(5)
			& 28393(9)\\
			CAS(20e,147o)\tnote{e}
			& 8007(19)
			& 10825(13)
			& 21252(13)
			& 29019(13)\\ \midrule
			MRCISD\tnote{f}
			&7633&10683&21507&29707\\
			\multicolumn{1}{l|}{Expt.\tnote{g}}
			& 7819
			& 10650
			& 21458
			& 29467 \\\bottomrule
		\end{tabular}
		\begin{tablenotes}
			\item[a] 20 electrons in $5p5d6s6p7s7p6d$ orbitals.
			\item[b] 20 electrons in $5p5d6s6p7s7p6d8s8p7d$ orbitals.
			\item[c] 20 electrons in $5p5d6s6p7s7p6d8s8p7d9s9p8d$ orbitals.
			\item[d] 20 electrons in $5p5d6s6p7s7p6d8s8p7d9s9p8d10s10p9d$ orbitals.
			\item[e] 20 electrons in $5p5d6s6p$ and all virtual orbitals.
			\item[f] Two-component large-core pseudopotential and core-polarization potential based MRCISD\cite{ccpVDZPP1}.
			\item[g] Experiments\cite{NIST1}.
		\end{tablenotes}
		\label{Pb_20e_orb}
	\end{threeparttable}
\end{table}

\begin{table}[!htp]
	\centering
	\caption{SA-iCISCF(21e,21o) orbital-based SOiCI energies (in cm$^{-1}$)
		for the low-lying states of \ce{Bi} relative to the ground state $^4S_{3/2}^{\circ}$.
		In parentheses are standard deviations of linear extrapolations}
	\begin{threeparttable}
		\begin{tabular}{l|llll}\toprule
			%			\multirow{3}{*}{active space}
			%			&\multicolumn{4}{c}{excited spinor states} \\ \cline{2-5}
			%			&\multicolumn{2}{c|}{(3/2,1/2)}
			%			&\multicolumn{2}{c}{(3/2,3/2)} \\\cline{2-5}
			%			& \multicolumn{1}{c}{$J=1$}
			%			& \multicolumn{1}{c|}{$J=2$}
			%			& \multicolumn{1}{c}{$J=2$}
			%			& \multicolumn{1}{c}{$J=0$} \\\toprule
			\multicolumn{1}{l|}{active space}
			& \multicolumn{1}{c}{$^2D_{3/2}^{\circ}$}
			& \multicolumn{1}{c}{$^2D_{5/2}^{\circ}$}
			& \multicolumn{1}{c}{$^2P_{1/2}^{\circ}$}
			& \multicolumn{1}{c}{$^2P_{3/2}^{\circ}$} \\ \toprule
			CAS(21e,21o)\tnote{a}
			& 11997(2)
			& 16344(2)
			& 23920(2)
			& 32114(2)\\
			CAS(21e,133o)\tnote{e}
			& 11121(35)
			& 15293(47)
			& 21382(46)
			& 33149(45)\\ \midrule
			\multicolumn{1}{l|}{Expt.\tnote{g}}
			& 11419
			& 15438
			& 21661
			& 33165 \\\bottomrule
		\end{tabular}
		\begin{tablenotes}
			\item[a] 21 electrons in $5p5d6s6p7s7p6d$ orbitals.
			\item[b] 21 electrons in $5p5d6s6p$ and all virtual orbitals.
			\item[c] Experiments\cite{NIST1}.
		\end{tablenotes}
		\label{Bi_21e_orb}
	\end{threeparttable}
\end{table}

%\begin{table}[!htp]
%	\centering
%	\caption{Energy levels (in cm$^{-1}$) of spinor states of \ce{I}, \ce{Tl} and \ce{Pb} with contracted and $p$-uncontracted ANO-RCC basis sets. All virtual orbitals are included in the SOiCI calculations.}
%	\begin{threeparttable}
%		\begin{tabular}{c|c|c|ccc|c} \toprule
%			atom
%			& configuration
%			& term
%			& $J$
%			& contracted
%			& $p$-uncontracted
%			& Expt. \\\toprule
%			\ce{I}\tnote{a}
%			& $5s^25p^5$
%			& $^2P^{\circ}$
%			& $1/2$
%			& 7416 & 7336 & 7603 \\\midrule
%			\ce{Tl}\tnote{b}
%			& $6s^26p^1$
%			& $^2P^{\circ}$
%			& $1/2$
%			& 7495 & 7377 & 7793 \\\midrule
%			\multirow{4}{*}{\ce{Pb}\tnote{c}}
%			& \multirow{4}{*}{$6s^26p^2$}
%			& \multirow{2}{*}{$(3/2,1/2)$}
%			& 1
%			& 8007  & 7871 & 7819\\
%			&&
%			& 2
%			& 10825 & 10539& 10650\\ \cline{3-7}
%			&&
%			\multirow{2}{*}{$(3/2,3/2)$}
%			& 2
%			& 21252 & 20817& 21458\\
%			&&
%			& 0
%			& 29019 & 28553& 29467\\\midrule
%		\end{tabular}
%	\begin{tablenotes}
%		\item[a] State-averaged iCISCF(17e,18o) orbitals with three $^2P$ scalar states averaged with equal weights.
%		\item[b] State-averaged iCISCF(19e,21o) orbitals with three $^2P$ scalar states averaged with equal weights.
%		\item[c] State-averaged iCISCF(20e,21o) orbitals with three $^3P$, five $^1D$ and one $^1S$ scalar states averaged with equal weights.
%	\end{tablenotes}
%	\end{threeparttable}
%\label{uncontracted}
%\end{table}

\begin{table}
	\centering
	\caption{Spin-orbit splitting (in cm$^{-1}$) of halogen atoms calculated by different methods.
In parentheses are standard deviations of linear extrapolations (cf. Table \ref{Detail_Data_Cl_Br_I_SOiCI})}
	\begin{threeparttable}
		%		\begin{tabular}{clllccc}\toprule
		\begin{tabular}{cllccc}\toprule	
			\multirow{1}{*}{atom}
			%			&\multirow{1}{*}{Hamiltonian}
			%			& \multicolumn{1}{c}{active space}
			& \multicolumn{1}{c}{SOiCI}
			& \multicolumn{1}{c}{SOSHCI\tnote{a}}
			& \multicolumn{1}{c}{EOM-CCSD(SOC)\tnote{b}}
			& \multicolumn{1}{c}{X2Cmmf-FSCCSD\tnote{c}}
			& Expt.\tnote{d}                              \\\toprule
			\multirow{1}{*}{\ce{F}}
			%			& (7e,87o)
			& 398(1)\tnote{e}
			& 399
			& 398
			& 421
			& \multirow{1}{*}{404}                       \\\midrule
			\ce{Cl}
			%			& (17e,100o)
%			& 871(8)\tnote{e}
            & 871(3)\tnote{f}
			& 866(1)
			& 876
			& 907
			& 882                       \\\midrule
			
			\multirow{1}{*}{\ce{Br}}
			%			& (17e,100o)
			& 3745(58)\tnote{g}
			& 3454(93)
			& 3649
			& 3723
			& 3685                      \\\midrule
			
			\ce{I}
			%			& (17e,121o)
			& 7416(38)\tnote{g}
			& 7487(62)
			& 7755
			& 7752
			& 7603                      \\\bottomrule
			
		\end{tabular}
		\begin{tablenotes}
			\item[a] Spin-orbit stochastic heat-bath CI\cite{HBCISOC2017}.
			\item[b] Ref.\cite{sf-X2C-EOM-SOC2017} (SOC included perturbatively in the CCSD step).
			\item[c] Ref.\cite{sf-X2C-EOM-SOC2017} (spinor-based Fock space CCSD).
			%			\item[d] Experiments\cite{NIST1}.
			\item[d] Experiments\cite{NIST1}.
			\item[e] CASSCF(7e,8o) orbitals.
			\item[f] CASSCF(13e,16o) orbitals.
			\item[g] iCISCF(17e,18o) orbitals.
		\end{tablenotes}
	\end{threeparttable}
	\label{Atom_Extrapolation}
\end{table}

\subsection{SOS of $p$-block Molecules}\label{Diatomics}
As a further test of SOiCI and iCISO, we consider the isovalent $^2\Pi$ diatomic molecules XO (X = \ce{N}, \ce{P}, \ce{As}, \ce{Sb}, \ce{Bi}) and XF (X = \ce{C}, \ce{Si}, \ce{Ge}, \ce{Sn}, \ce{Pb}),
at their experimental bond lengths (1.15077~\AA~ for NO, 1.476~\AA~
for PO, 1.6236~\AA~ for AsO, 1.825~\AA~ for SbO, 1.934~\AA~ for
BiO, 1.272~\AA~ for CF, 1.601~\AA~ for SiF, 1.745~\AA~ for GeF,
1.944~\AA~ for SnF, and 2.0575~\AA~ for PbF)\cite{NIST2}.

Following the previous atomic calculations, the minimal active space CAS(1e,2o) was first considered here.
As can be seen from Table \ref{SmallActiveSpaceResult}, the results are pretty off. However, unlike the atomic calculations,
larger active spaces cannot readily be chosen solely by means of orbital energies. Instead, the iCAS (imposed automatic selection of CAS)
approach\cite{iCAS} can be adopted here, which employs valence atomic orbitals (VAO) as probe to select precisely the same number of
guess orbitals for CASSCF/iCISCF. Moreover, iCAS imposes the matching of the doubly occupied, active, and virtual subspaces between two adjacent iterations,
so as to render the converged CASSCF orbitals as close to the guest orbitals as possible.
Specifically, the
%\textcolor{red}{
$2s,2p,3s,3p,3d$ atomic shells (13 VAOs) for the second row atoms,
%}
%\textcolor{green}{$2s,2p,3s,3p$ atomic shells (8 VAOs) for the second row atoms},
%\textcolor{red}{
the $2s,2p,3s,3p,3d,4s,4p$ atomic shells (17 VAOs) for the third row atoms,
%}
%\textcolor{green}{the $2p,3s,3p,3d,4s,4p$ atomic shells (16 VAOs) for the third row atoms},
and the $(n-1)d,ns,np,(n+1)s,(n+1)p,(n+1)d$ atomic shells (18 VAOs) for the $n$-th row atoms ($n>3$). Such VAOs can readily be obtained
from sf-X2C-HF calculations of spherical, unpolarized atomic calculations. The corresponding active spaces are then
%\textcolor{red}{
	CAS(11e,26o), CAS(19e,30o), and CAS(21e,31o)
%}
%\textcolor{green}{CAS(11e,16o), CAS(17e,24o), and CAS(21e,26o)}
for the 2-2, 3-2, and $n$-2 ($n>3)$ types of diatomic molecules, respectively (NB: $n$-$m$ is short for $nth$ row-$m$th row).
The so-calculated results are given in Table \ref{EOMCC}. It can be seen that both iCISO and SOiCI work very well for these systems.
The former is of course due to the fact that the SOC is quenched substantially as compared to the free atoms. The results are even better than those
by the EOMEA-CCSD (equation-of-motion electron affinity coupled-cluster singles and doubles)\cite{Diatomic_CCSD}  with all electrons correlated
in the uncontracted ANO-RCC basis sets.

\begin{table}
	\centering
	\caption{Spin-orbit splitting (in cm$^{-1}$) of \ce{XO} and \ce{XF} calculated by SOiCI with the minimal active space CAS(1e,2o)}
	\begin{threeparttable}
		\begin{tabular}{ccccccccc}\toprule
			\multirow{1}{*}{molecule}
			& \multirow{1}{*}{CAS(1e,2o)}
			& \multirow{1}{*}{Expt.\tnote{a}}
			& \multirow{1}{*}{molecule}& \multirow{1}{*}{CAS(1e,2o)}&\multirow{1}{*}{Expt.\tnote{a}} \\ \toprule
			NO       & 125          & 120  &CF       & 64           & 77 \\
			PO       & 239          & 224  &SiF      & 135          & 156\\
			AsO      & 1234         & 1026 &GeF      & 976          & 934\\
			SbO      & 2614         & 2272 &SnF      & 2256         & 2317\\
			BiO      & 8655         & 7089 &PbF      & 7806         & 8264\\\midrule
			\multicolumn{6}{l}{percentage mean absolute error:  $11.1\pm 6.4$ }       \\ \bottomrule
		\end{tabular}
		\begin{tablenotes}
			\item[a] Experiments\cite{NIST2}.
		\end{tablenotes}
	\end{threeparttable}
	\label{SmallActiveSpaceResult}
\end{table}

\begin{table}
	% CASSCF
	\small
	\centering
	\caption{Spin-orbit splitting of XO and XF calculated by different methods
		%		\textcolor{red}{CHANGE DATA!}
	}
	\begin{threeparttable}
		\begin{tabular}{cccccc}\toprule
			molecule             & SOiCI\tnote{a}   & iCISO\tnote{a}   & EOMEA(SO)-CCSD\tnote{b} & EOMEA-CCSD(SO)\tnote{c} & Expt.\tnote{d} \\\toprule
			NO                   & 120              & 119              & 122                     & 122                     & 120  \\
			PO                   & 230              & 232              & 230                     & 218                     & 224  \\
			AsO                  & 958              & 954              & 1089                    & 967                     & 1026 \\
			SbO                  & 2326             & 2305             & 2518                    & 2142                    & 2272 \\
			BiO                  & 7542             & 7167             & 7598                    & 6171                    & 7089 \\\midrule
			CF                   & 75               & 75               & 75                      & 78                      & 77   \\
			SiF                  & 166              & 166              & 163                     & 158                     & 156  \\
			GeF                  & 887              & 882              & 949                     & 907                     & 934  \\
			SnF                  & 2300             & 2212             & 2288                    & 2176                    & 2317 \\
			PbF                  & 8586             & 7561             & 7150                    & 6777                    & 8264 \\\midrule
			PMAE\tnote{e}        & 3.7$\pm2.0$      & 4.1$\pm 2.3$      & 5.3$\pm3.3$             & 5.8$\pm3.9$             &      \\
			\bottomrule
		\end{tabular}
		\begin{tablenotes}
			\item[a] iCISCF orbitals with contracted ANO-RCC basis ($C_{\mathrm{min}}=5\times 10^{-6}$).
%			\item[b] iCISCF orbitals with $p$-decontracted ANO-RCC basis ($C_{\mathrm{min}}=5\times 10^{-6}$).
			\item[b] Ref.\cite{Diatomic_CCSD} (uncontracted ANO-RCC basis; SOC included perturbatively in the EOM step).
			\item[c] Ref.\cite{Diatomic_CCSD} (uncontracted ANO-RCC basis; SOC included perturbatively in the CCSD step).
			\item[d] Experiments\cite{NIST2}.
			\item[e] Percentage mean absolute error.
		\end{tablenotes}
	\end{threeparttable}
	\label{EOMCC}
\end{table}

\section{Conclusion}\label{Conclusion}
Two, two-step relativistic approaches, SOiCI and iCISO, have been proposed to treat scalar relativity and SOC separately.
The former amounts to treating SOC and correlation on an equal footing, whereas the latter
is rooted in quasi-degenerate perturbation theory and is hence applicable only when spin-dependent orbit relaxation
is not significant or when SOC is quenched to a large extent by ligand fields in the case of heavy atoms.
The use of both double group and time reversal adapted many-electron basis facilitates greatly not only
the computation but also the analysis of spinor wave functions. Although only pilot calculations have been performed
to elucidate mainly the conceptual aspects on the interplay between SOC and correlation, there is no doubt that the methods can widely be applied to investigate the SOC
in general open-shell systems containing heavy elements.

\section*{Acknowledgment}
This work was supported by the National Natural Science Foundation of China (Grant Nos. 21833001 and 21973054),
Mountain Tai Climbing Program of Shandong Province, and Key-Area Research and Development Program of Guangdong Province (Grant No. 2020B0101350001).

\section*{Data Availability Statement}
The data that supports the findings of this study is available within the article.

\section*{Postscript}
This work is dedicated to the commemoration of Prof. Dr. Enrico Clementi, a good friend of one of the present authors (WL), both personally and scientifically. 
His expertise and ideas in relativistic quantum chemistry and computational chemistry had great influence on WL's early research.

\clearpage
\newpage
\appendix

\section{Double Group Symmetry}\label{AppDG}

To complete the discussion of double group symmetry presented in the main text, the spin rotation matrices $\mathbf{S}^{(g)}$ in Eq. \eqref{VSmat}
should further be specified. Instead of the parametrization \eqref{SU2def}, the following
parametrization
\begin{equation}
\mathbf{R}^{\frac{1}{2}}(\alpha,\beta,\gamma) =
\begin{pmatrix}
e^{-\mathbbm{i}\frac{\gamma+\alpha}{2}}\cos\frac{\beta}{2}
&-e^{\mathbbm{i}\frac{\gamma-\alpha}{2}}\sin\frac{\beta}{2} \\
e^{-\mathbbm{i}\frac{\gamma-\alpha}{2}}\sin\frac{\beta}{2}
&e^{\mathbbm{i}\frac{\gamma+\alpha}{2}}\cos\frac{\beta}{2} \\
\end{pmatrix}
=\mathbf{R}^{\frac{1}{2}}(z,\alpha)
\mathbf{R}^{\frac{1}{2}}(y,\beta))
\mathbf{R}^{\frac{1}{2}}(z,\gamma)
\end{equation}
in terms of the Euler angles $\alpha, \beta, \gamma$ is more convenient for the construction of $\mathbf{S}^{(g)}$. Specifically,
\begin{eqnarray}
S^{(g)}_{M'M}=\langle SM'|\mathbf{R}^{\frac{1}{2}}(\alpha,\beta,\gamma)|SM\rangle = e^{-\mathbbm{i}M'\alpha} d^S_{M'M}(\beta) e^{-\mathbbm{i}M \gamma}, \label{SU2_Mat}
\end{eqnarray}
where
\begin{eqnarray}
d_{M^{\prime} M}^{S}(\beta)&=&
\left[\left(S+M^{\prime}\right) !\left(S-M^{\prime}\right) !(S+M) !(S-M) !\right]^{1 / 2} \nonumber\\
&\times& \sum_{s} \frac{(-1)^{M^{\prime}-M+s}\left(\cos \frac{\beta}{2}\right)^{2 S+M-M^{\prime}-2 s}\left(\sin \frac{\beta}{2}\right)^{M^{\prime}-M+2 s}}{(S+M-s) ! s !\left(M^{\prime}-M+s\right) !\left(S-M^{\prime}-s\right) !}.\label{dmat}
\end{eqnarray}
The calculation is simplified greatly by noticing that the operations of $D_2^\ast$ can mix only two spin functions $|SM\rangle$ and $|S-M\rangle$.
The corresponding matrix elements $S^{(g)}_{M'M}$ are documented in Tables \ref{d2even} and \ref{d2odd} for even and odd numbers of electrons, respectively.
The same spin rotation matrices apply also to $D_{2h}^\ast$, because $g\in D_{2h}^\ast$ can be written as $g=g'c$ ($g'\in D_{2}^\ast$, $c\in C_i=\{E,i\}$) and
the spin rotation matrices of $C_i$ are simply unit matrices. The spin rotation matrices of other subgroups can be
read from the tables by looking up the corresponding symmetry operations. Further discussions on the spin rotation matrices can be found from Sec. \ref{AppTR}.

\begin{table}
	\centering
	\caption{Reducible spin rotation matrices $\mathbf{S}^{(g)}$ of $D_2^\ast$ in the basis $(|SM\rangle,|S-M\rangle)$ with $S$ being an integer}
	\begin{threeparttable}[!htp]
		\tiny
		\begin{tabular}{c|c|c|c|c|c|c|c|c}\toprule
			$(|SM\rangle,|S-M\rangle)$
			& $E$
			& $\bar{E}$
			& $-\mathbbm{i}\sigma_x$
			& $-\mathbbm{i}\sigma_x\bar{E}$
			& $-\mathbbm{i}\sigma_y$
			& $-\mathbbm{i}\sigma_y\bar{E}$
			& $-\mathbbm{i}\sigma_z$
			& $-\mathbbm{i}\sigma_z\bar{E}$\\ \toprule
			\multicolumn{9}{c}{$S$ odd} \\ \cline{1-9}
			$M=0$
			&1&1&-1&-1&-1&-1&1&1\\\cline{1-9}	
			$M$ odd&
			$\begin{pmatrix}
			1 & 0 \\
			0 & 1 \\
			\end{pmatrix}$
			&
			$\begin{pmatrix}
			1 & 0 \\
			0 & 1 \\
			\end{pmatrix}$
			&
			$\begin{pmatrix}
			0  & -1  \\
			-1 &  0  \\
			\end{pmatrix}$
			&
			$\begin{pmatrix}
			0 & -1 \\
			-1 & 0 \\
			\end{pmatrix}$
			&
			$\begin{pmatrix}
			0 & 1 \\
			1 & 0 \\
			\end{pmatrix}$
			&
			$\begin{pmatrix}
			0 & 1 \\
			1 & 0 \\
			\end{pmatrix}$
			&
			$\begin{pmatrix}
			-1 & 0 \\
			0 & -1 \\
			\end{pmatrix}$
			&
			$\begin{pmatrix}
			-1 & 0 \\
			0 & -1 \\
			\end{pmatrix}$ \\\cline{1-9}
			$M$ even&
			$\begin{pmatrix}
			1 & 0 \\
			0 & 1 \\
			\end{pmatrix}$
			&
			$\begin{pmatrix}
			1 & 0 \\
			0 & 1 \\
			\end{pmatrix}$
			&
			$\begin{pmatrix}
			0 & -1  \\
			-1 &  0  \\
			\end{pmatrix}$
			&
			$\begin{pmatrix}
			0 & -1 \\
			-1 & 0 \\
			\end{pmatrix}$
			&
			$\begin{pmatrix}
			0 & -1 \\
			-1 & 0 \\
			\end{pmatrix}$
			&
			$\begin{pmatrix}
			0 & -1 \\
			-1 & 0 \\
			\end{pmatrix}$
			&
			$\begin{pmatrix}
			1 & 0 \\
			0 & 1 \\
			\end{pmatrix}$
			&
			$\begin{pmatrix}
			1 & 0 \\
			0 & 1 \\
			\end{pmatrix}$ \\\cline{1-9}
			
			\multicolumn{9}{c}{$S$ even} \\ \cline{1-9}
			$M=0$
			&1&1&1&1&1&1&1&1\\\cline{1-9}	
			$M$ odd&
			$\begin{pmatrix}
			1 & 0 \\
			0 & 1 \\
			\end{pmatrix}$
			&
			$\begin{pmatrix}
			1 & 0 \\
			0 & 1 \\
			\end{pmatrix}$
			&
			$\begin{pmatrix}
			0  & 1  \\
			1 &  0  \\
			\end{pmatrix}$
			&
			$\begin{pmatrix}
			0 & 1 \\
			1 & 0 \\
			\end{pmatrix}$
			&
			$\begin{pmatrix}
			0 & -1 \\
			-1 & 0 \\
			\end{pmatrix}$
			&
			$\begin{pmatrix}
			0 & -1 \\
			-1 & 0 \\
			\end{pmatrix}$
			&
			$\begin{pmatrix}
			-1 & 0 \\
			0 & -1 \\
			\end{pmatrix}$
			&
			$\begin{pmatrix}
			-1 & 0 \\
			0 & -1 \\
			\end{pmatrix}$ \\\cline{1-9}
			$M$ even&
			$\begin{pmatrix}
			1 & 0 \\
			0 & 1 \\
			\end{pmatrix}$
			&
			$\begin{pmatrix}
			1 & 0 \\
			0 & 1 \\
			\end{pmatrix}$
			&
			$\begin{pmatrix}
			0 &  1  \\
			1 &  0  \\
			\end{pmatrix}$
			&
			$\begin{pmatrix}
			0 &  1 \\
			1 &  0 \\
			\end{pmatrix}$
			&
			$\begin{pmatrix}
			0 & 1 \\
			1 & 0 \\
			\end{pmatrix}$
			&
			$\begin{pmatrix}
			0 & 1 \\
			1 & 0 \\
			\end{pmatrix}$
			&
			$\begin{pmatrix}
			1 & 0 \\
			0 & 1 \\
			\end{pmatrix}$
			&
			$\begin{pmatrix}
			1 & 0 \\
			0 & 1 \\
			\end{pmatrix}$ \\\cline{1-9}
		\end{tabular}
	\end{threeparttable}
	\label{d2even}
\end{table}

\begin{table}
	\centering
	\caption{Irreducible spin rotation matrices $\mathbf{S}^{(g)}$ of $D_2^\ast$ in the basis $(|SM\rangle,|S-M\rangle)$ with $S$ being a half-integer}
	\begin{threeparttable}
		\tiny
		\begin{tabular}{c|c|c|c|c|c|c|c|c}\toprule
			$(|SM\rangle,|S-M\rangle)$
			& $E$
			& $\bar{E}$
			& $-\mathbbm{i}\sigma_x$
			& $-\mathbbm{i}\sigma_x\bar{E}$
			& $-\mathbbm{i}\sigma_y$
			& $-\mathbbm{i}\sigma_y\bar{E}$
			& $-\mathbbm{i}\sigma_z$
			& $-\mathbbm{i}\sigma_z\bar{E}$\\ \toprule
			\multicolumn{9}{c}{$S=p+\frac{1}{2}$, $p$ odd} \\ \cline{1-9}	
			$M=q+\frac{1}{2}$, $q$ odd&
			$\begin{pmatrix}
			1 & 0 \\
			0 & 1 \\
			\end{pmatrix}$
			&
			$\begin{pmatrix}
			-1 & 0 \\
			0 & -1 \\
			\end{pmatrix}$
			&
			$\begin{pmatrix}
			0  &  \mathbbm{i}  \\
			\mathbbm{i} &  0  \\
			\end{pmatrix}$
			&
			$\begin{pmatrix}
			0  & -\mathbbm{i} \\
			-\mathbbm{i} & 0 \\
			\end{pmatrix}$
			&
			$\begin{pmatrix}
			0 & -1 \\
			1 & 0 \\
			\end{pmatrix}$
			&
			$\begin{pmatrix}
			0 & 1 \\
			-1 & 0 \\
			\end{pmatrix}$
			&
			$\begin{pmatrix}
			\mathbbm{i} & 0 \\
			0 & -\mathbbm{i}\\
			\end{pmatrix}$
			&
			$\begin{pmatrix}
			-\mathbbm{i} & 0 \\
			0  & \mathbbm{i} \\
			\end{pmatrix}$ \\\cline{1-9}
			$M=q+\frac{1}{2}$, $q$ even&
			$\begin{pmatrix}
			1 & 0 \\
			0 & 1 \\
			\end{pmatrix}$
			&
			$\begin{pmatrix}
			-1 & 0 \\
			0 & -1 \\
			\end{pmatrix}$
			&
			$\begin{pmatrix}
			0 & \mathbbm{i}  \\
			\mathbbm{i} & 0  \\
			\end{pmatrix}$
			&
			$\begin{pmatrix}
			0 & -\mathbbm{i} \\
			-\mathbbm{i} & 0 \\
			\end{pmatrix}$
			&
			$\begin{pmatrix}
			0 & 1 \\
			-1 & 0 \\
			\end{pmatrix}$
			&
			$\begin{pmatrix}
			0 & -1 \\
			1 & 0 \\
			\end{pmatrix}$
			&
			$\begin{pmatrix}
			-\mathbbm{i} & 0 \\
			0  & \mathbbm{i} \\
			\end{pmatrix}$
			&
			$\begin{pmatrix}
			\mathbbm{i} & 0  \\
			0 & -\mathbbm{i} \\
			\end{pmatrix}$ \\\cline{1-9}
			
			\multicolumn{9}{c}{$S=p+\frac{1}{2}$, $p$ even} \\ \cline{1-9}
			$M=q+\frac{1}{2}$, $q$ odd&
			$\begin{pmatrix}
			1 & 0 \\
			0 & 1 \\
			\end{pmatrix}$
			&
			$\begin{pmatrix}
			-1 & 0 \\
			0 & -1 \\
			\end{pmatrix}$
			&
			$\begin{pmatrix}
			0  & -\mathbbm{i}  \\
			-\mathbbm{i} &  0  \\
			\end{pmatrix}$
			&
			$\begin{pmatrix}
			0 & \mathbbm{i} \\
			\mathbbm{i} & 0 \\
			\end{pmatrix}$
			&
			$\begin{pmatrix}
			0 & 1 \\
			-1 & 0 \\
			\end{pmatrix}$
			&
			$\begin{pmatrix}
			0 & -1 \\
			1 & 0 \\
			\end{pmatrix}$
			&
			$\begin{pmatrix}
			\mathbbm{i} & 0  \\
			0 & -\mathbbm{i} \\
			\end{pmatrix}$
			&
			$\begin{pmatrix}
			-\mathbbm{i} & 0 \\
			0 &  \mathbbm{i} \\
			\end{pmatrix}$ \\\cline{1-9}
			$M=q+\frac{1}{2}$, $q$ even&
			$\begin{pmatrix}
			1 & 0 \\
			0 & 1 \\
			\end{pmatrix}$
			&
			$\begin{pmatrix}
			-1 & 0 \\
			0 & -1 \\
			\end{pmatrix}$
			&
			$\begin{pmatrix}
			0 &  -\mathbbm{i}  \\
			-\mathbbm{i} &  0  \\
			\end{pmatrix}$
			&
			$\begin{pmatrix}
			0 &  \mathbbm{i} \\
			\mathbbm{i} &  0 \\
			\end{pmatrix}$
			&
			$\begin{pmatrix}
			0 & -1 \\
			1  & 0 \\
			\end{pmatrix}$
			&
			$\begin{pmatrix}
			0  & 1 \\
			-1 & 0 \\
			\end{pmatrix}$
			&
			$\begin{pmatrix}
			-\mathbbm{i} & 0 \\
			0  & \mathbbm{i} \\
			\end{pmatrix}$
			&
			$\begin{pmatrix}
			\mathbbm{i} & 0  \\
			0 & -\mathbbm{i} \\
			\end{pmatrix}$ \\\cline{1-9}
		\end{tabular}
	\end{threeparttable}
	\label{d2odd}
\end{table}

% all the case

\section{Time reversal symmetry}\label{AppTR}
Double group adapted basis functions do not form naturally Kramers pairs, which can only be achieved by
further incorporating properly the time reversal symmetry. As shown\cite{Symm2009} before, this can actually be
done by a simple unitary transformation of the symmetrized functions.

A set of functions $\{|\tau\mu i\rangle,i=1,\cdots,n_\mu\}$ spanning an irrep $\mu$ of double group $G^\ast$
will transform among themselves according to
\begin{equation}
g | \tau\mu i\rangle = \sum_j |\tau\mu j\rangle D_{ji}^{(\mu)}(g),
\end{equation}
where $\mathbf{D}^{(\mu)}(g)$ is an $n_{\mu}$-dimensional unitary matrix corresponding to operation $g$.
The index $i$ in $|\tau\mu i\rangle$ indicates that the function belongs to the $i$th column of irrep $\mu$, whereas $\tau$ serves to distinguish functions of the same $\mu$ and $i$.
The Hamiltonian matrix over such symmetrized basis functions are block diagonal, i.e.,
\begin{equation}
\langle \tau \mu i |H|\sigma\nu j \rangle = \delta_{\mu\nu}\delta_{ij} \langle \tau \mu i |H|\sigma\nu i \rangle
= \delta_{\mu\nu}\delta_{ij}\langle \tau\mu \|H\|\sigma\nu \rangle, \label{reducedH}
\end{equation}
where $\langle \tau\mu \|H\|\sigma\nu \rangle$ are the so-called reduced matrix elements. Moreover, the fact
that the time reversal operation $\mathcal{T}$ commutes with the Hamiltonian $H$ imposes the following relations between the Hamiltonian matrix elements
\begin{eqnarray}
	\langle \mathcal{T} \psi_p | H | \mathcal{T} \psi_q \rangle&=& \langle \psi_p | H | \psi_q \rangle ^\ast, \label{TH1}\\
	\langle \mathcal{T} \psi_p | H |  \psi_q \rangle &=& \pm\langle \psi_p | H | \mathcal{T}\psi_q \rangle ^\ast, \quad
	\mathcal{T}^2\psi_p=\pm\psi_p, \label{TH2}
\end{eqnarray}
where the positive and negative signs apply to boson and fermion types of functions, respectively.
In particular, it is easy to see from the action of $\mathcal{T}$ on the spin functions $\{|SM\rangle\}_{M=-S}^S$,
\begin{eqnarray}
\mathcal{T}|SM\rangle = (-1)^{S-M}|S-M\rangle,\label{Time_SM}
\end{eqnarray}
that the spin functions are either boson (if $S$ is an integer for an even number of electrons) or fermion functions (if $S$ is a half-integer
for an odd number of electrons).

Since the irreducible representation matrices for the spaces spanned by $\{|\tau\mu i\rangle\}$ and $\{\mathcal{T}|\tau\mu i\rangle\}$
are related simply by complex conjugation, viz.,
\begin{equation}
g \mathcal{T}| \tau\mu i\rangle = \mathcal{T}g | \tau\mu i\rangle= \sum_j D_{ji}^{(\mu)\ast}(g) \mathcal{T}|\tau\mu j\rangle,
\end{equation}
they can be classified into three Frobenius-Schur classes:
\begin{enumerate}[(a)]
	\item $\mathbf{D}^{(\mu)}(g)$ are equivalent to $\mathbf{D}^{(\mu)^\ast}(g)$ and can be made real-valued by a suitable unitary transformation.
	\item $\mathbf{D}^{(\mu)}(g)$ are inequivalent to $\mathbf{D}^{(\mu)^\ast}(g)$.
	\item $\mathbf{D}^{(\mu)}(g)$ are equivalent to $\mathbf{D}^{(\mu)^\ast}(g)$ but cannot be made real-valued by any unitary transformation.
\end{enumerate}
For convenience, the Frobenius-Schur classes for the irreps of $D_{2h}^\ast$ and its subgroups are documented in Table \ref{FSClass}.
It is clear that all boson irreps spanned by
the spin functions of an even number of electrons all belong to class (a). Without loss of generality,
$|SM\rangle$ and $|S-M\rangle$ can in this case be recombined to form two eigenvectors, $|SM+\rangle$ and $|SM-\rangle$, of $\mathcal{T}$ with eigenvalue 1, viz.,
\begin{equation}
(|SM+\rangle, |SM-\rangle)=(|SM\rangle, |S-M\rangle)\begin{pmatrix} \frac{\mathbbm{i}^S(-1)^M}{\sqrt{2+2\delta_{M,0}}} & \frac{\mathbbm{i}^{S+1}(-1)^{M+1}}{\sqrt{2}}\\
 \frac{\mathbbm{i}^S}{\sqrt{2+2\delta_{M,0}}}&\frac{\mathbbm{i}^{S+1}}{\sqrt{2}}
 \end{pmatrix},\label{SM+-}
\end{equation}
where $M\ge 0$ for $|SM+\rangle$ and $M>0$ for $|SM-\rangle$.
%\begin{subequations}\label{SM+-}
%	\begin{equation}
%	|SM-\rangle = \mathbbm{i}^{S+1}/\sqrt{2}\left[|S,-M\rangle-(-1)^M|S,M\rangle\right],
%	\end{equation}
%	\begin{equation}
%	|SM+\rangle = \mathbbm{i}^S\frac{1}{\sqrt{2+2\delta_{M,0}}}\left[|S,-M\rangle+(-1)^M|S,M\rangle\right],
%	\end{equation}
%\end{subequations}
Both $|SM+\rangle$ and $|SM-\rangle$ can serve as a basis for the 1D irreps of $D_2^\ast$. As such, the corresponding
spin rotation matrices $\mathbf{S}^{(g)}$ of $D_2^\ast$ (obtained from the above transformation of those documented in Table \ref{d2even})
become irreducible, see Table \ref{d2_even_adapted}. The associations of $|SM\pm\rangle$
with the irreps of $D_{2h}^\ast$ and its subgroups are further given in Table \ref{species_even}.
It is trivial to see that the Hamiltonian matrix elements in this basis and hence the CI vectors are real-valued.
%\begin{equation}
%\mathcal{T}H|\mu\rangle =\sum_{\nu}|\nu\rangle \langle\nu |H|\mu\rangle^\ast= H|\mu\rangle= \sum_{\nu}|\nu\rangle \langle\nu |H|\mu\rangle, \quad \forall \mu \in\{|J\mathbf{d}s\rangle\}.
%\end{equation}
That is, time reversal symmetry itself will reduce the computational cost by a factor of two as compared with the complex algebra without time reversal symmetry.
Further combined with spatial symmetry, a factor of $2|G|=|G^*|$ will be gained in efficiency.

\begin{table}
	\centering
	\caption{The Frobenius-Schur classes (a,b,c) of $D_{2h}^\ast$ and its subgroups. In parentheses are
the orders of the groups}
\begin{threeparttable}
	\begin{tabular}{ccc}\toprule
	 boson& fermion & point groups \\\toprule
		a & a       & $C_1^\ast$  (2), $C_i^\ast$ (4)\\
		a & b       & $C_2^\ast$ (4), $C_s^\ast$ (4), $C_{2h}^\ast$ (8) \\
		a & c       & $C_{2v}^\ast$ (8), $D_2^\ast$ (8), $D_{2h}^\ast$ (16)\\\bottomrule
	\end{tabular}
\end{threeparttable}
\label{FSClass}
\end{table}

\begin{table}
	\centering
	\caption{Irreducible spin rotation representations $\mathbf{S}^{(g)}$ of $D_2^\ast$ in the basis $|SM\pm\rangle$ \eqref{SM+-}$^\ast$.}
	\small
	\begin{threeparttable}
		\begin{tabular}{c|c|c|c|c|c|c|c|c}\toprule
			%		$\{|SM\rangle,|S-M\rangle\}$
			& $E$
			& $\bar{E}$
			& $-\mathbbm{i}\sigma_x$
			& $-\mathbbm{i}\sigma_x\bar{E}$
			& $-\mathbbm{i}\sigma_y$
			& $-\mathbbm{i}\sigma_y\bar{E}$
			& $-\mathbbm{i}\sigma_z$
			& $-\mathbbm{i}\sigma_z\bar{E}$\\ \toprule
			\multicolumn{9}{c}{$S$ odd} \\ \cline{1-9}
			%		$m=0$
			%		&1&1&-1&-1&-1&-1&1&1\\\cline{1-9}	
			$|SM+\rangle$, $M$ odd
			&1&1&1&1&-1&-1&-1&-1\\\cline{1-9}
			$|SM+\rangle$, $M$ even
			&1&1&-1&-1&-1&-1&1&1\\\cline{1-9}
			$|SM-\rangle$, $M$ odd
			&1&1&-1&-1&1&1&-1&-1\\\cline{1-9}
			$|SM-\rangle$, $M$ even
			&1&1&1&1&1&1&1&1\\\cline{1-9}
			\multicolumn{9}{c}{$S$ even} \\ \cline{1-9}
			%		$m=0$
			%		&1&1&1&1&1&1&1&1\\\cline{1-9}	
			$|SM+\rangle$, $M$ odd
			&1&1&-1&-1&1&1&-1&-1\\\cline{1-9}
			$|SM+\rangle$, $M$ even
			&1&1&1&1&1&1&1&1\\\cline{1-9}
			$|SM-\rangle$, $M$ odd
			&1&1&1&1&-1&-1&-1&-1\\\cline{1-9}
			$|SM-\rangle$, $M$ even
			&1&1&-1&-1&-1&-1&1&1\\\cline{1-9}
		\end{tabular}
\begin{tablenotes}
\item [*]$M\ge 0$ for $|SM+\rangle$ and $M>0$ for $|SM-\rangle$.
\end{tablenotes}
	\end{threeparttable}
	\label{d2_even_adapted}
\end{table}

%\begin{table}
%	\centering
%	\caption{Character table of $D_{2}$}
%	\begin{threeparttable}	
%		\begin{tabular}{lcccc}\toprule
%			& $E$ & $C_{2x}$ & $C_{2y}$ & $C_{2z}$ \\\toprule
%			$A_1$&1&1&1&1\\\midrule
%			$B_1$&1&-1&-1&1\\\midrule
%			$B_2$&1&-1&1&-1\\\midrule
%			$B_3$&1&1&-1&-1\\\midrule
%            $E_{1/2}$&2&0&0&0\\\midrule
%		\end{tabular}
%	\end{threeparttable}
%	\label{d2_character}
%\end{table}

\begin{table}
	\centering
	\caption{Symmetrized spin-only boson functions of $D_{2h}^\ast$ and its subgroups$^\ast$
	}
	\begin{threeparttable}
		\begin{tabular}{c|ccccccc}\toprule
			& $D_{2h}^\ast$ & $D_2^\ast$ & $C_{2h}^\ast$
			& $C_{2v}^\ast$ & $C_s^\ast$ & $C_i^\ast$ & $C_2^\ast$\\\toprule
			\multicolumn{8}{c}{$S$ odd} \\\toprule
			$|SM+\rangle$, $M$ odd
			&$B_{3g}$&$B_3$&$B_g$&$B_2$&$A''$&$A_g$&$B$\\\cline{1-8}
			$|SM+\rangle$, $M$ even
			&$B_{1g}$&$B_1$&$A_g$&$A_2$&$A'$&$A_g$&$A$\\\cline{1-8}
			$|SM-\rangle$, $M$ odd
			&$B_{2g}$&$B_2$&$B_g$&$B_1$&$A''$&$A_g$&$B$\\\cline{1-8}
			$|SM-\rangle$, $M$ even
			&$A_{g}$&$A_1$&$A_g$&$A_1$&$A'$&$A_g$&$A$\\\cline{1-8}
			\multicolumn{8}{c}{$S$ even} \\\toprule
			$|SM+\rangle$, $M$ odd
			&$B_{2g}$&$B_2$&$B_g$&$B_1$&$A''$&$A_g$&$B$\\\cline{1-8}
			$|SM+\rangle$, $M$ even
			&$A_{g}$&$A_1$&$A_g$&$A_1$&$A'$&$A_g$&$A$\\\cline{1-8}
			$|SM-\rangle$, $M$ odd
			&$B_{3g}$&$B_3$&$B_g$&$B_2$&$A''$&$A_g$&$B$\\\cline{1-8}
			$|SM-\rangle$, $M$ even
			&$B_{1g}$&$B_1$&$A_g$&$A_2$&$A'$&$A_g$&$A$\\\cline{1-8}
		\end{tabular}
\begin{tablenotes}
\item [*]$M\ge 0$ for $|SM+\rangle$ and $M>0$ for $|SM-\rangle$.
\end{tablenotes}
	\end{threeparttable}
	\label{species_even}
\end{table}

As for the case of an odd number of electrons, the fermion irreps of $D_{2h}^\ast$ and its subgroups may belong to class (a), (b) or (c) (cf. Table \ref{FSClass}).
Class (a) includes $C_1^\ast$ and $C_i^\ast$. In this case, every spin function $|SM\rangle$ serves as the basis of a 1D irrep.
So is every component $|J\mathbf{d}M\rangle$ of CSF $|J\mathbf{d}\rangle$, see Table \ref{TimeReversalCi}. In particular, $|J\mathbf{d}M\rangle$ and $\mathcal{T}|J\mathbf{d}M\rangle$ are
already Kramers paired, such that the Hamiltonian matrix has the following (quaternion) structure
	\begin{equation}
	(|J\mathbf{d}M\rangle,\mathcal{T}|J\mathbf{d}M\rangle)^\dagger H (|J\mathbf{d}M\rangle,\mathcal{T}|J\mathbf{d}M\rangle) =
	\begin{bmatrix}
	A & B \\
	-B^\ast & A^\ast
	\end{bmatrix},\label{HCase1}
	\end{equation}
in view of Eqs. \eqref{TH1} and \eqref{TH2}. This structure allows a reduction of the memory footprint by a factor of two. However,
since the computational cost is dominated heavily by the spin-free two-body terms, there is no much gain in efficiency in the evaluation of the Hamiltonian matrix elements.
On the other hand, although both $\{|J\mathbf{d}M\rangle\}$ and $\{\mathcal{T}|J\mathbf{d}M\rangle\}$ are needed to expand the CI eigenvectors,
the number of states is halved due to the Kramers degeneracy between states $|\Psi_k\rangle$ and $\mathcal{T}|\Psi_k\rangle$.
As such, the combined use of spatial and time reversal symmetries leads to
an overall reduction of the computational cost by a factor of two ($=2|C_1|=|C_1^\ast|$) and four ($=2|C_i|=|C_i^\ast|$) for $C_1^\ast$ and $C_i^\ast$, respectively.

As can be seen from Table \ref{FSClass}, the fermion irreps of $C_2^\ast$, $C_s^\ast$, and $C_{2h}^\ast$ belong to class (b).
In this case, every spin function $|SM\rangle$ also serves as the basis of a 1D irrep.
So is every component $|J\mathbf{d}M\rangle$ of CSF $|J\mathbf{d}\rangle$, see Tables \ref{TimeReversalC2}, \ref{TimeReversalCs}, and \ref{TimeReversalC2h} for $C_2^\ast$, $C_s^\ast$, and $C_{2h}^\ast$, respectively.
Furthermore, if $\{|J\mathbf{d}M\rangle\}$ span irrep $\mu$, $\{\mathcal{T}|J\mathbf{d}M\rangle\}$ will span a different irrep $\mu^\ast$. In view of Eq. \eqref{reducedH},
the Hamiltonian matrix has the following structure
\begin{equation}
(|J\mathbf{d}M\rangle,\mathcal{T}|J\mathbf{d}M\rangle)^\dagger H (|J\mathbf{d}M\rangle,\mathcal{T}|J\mathbf{d}M\rangle) =
\begin{bmatrix}
A & 0 \\
0 & A^\ast
\end{bmatrix}, \label{HCase2}
\end{equation}
where $A$ is in general complex.
Since $A$ and $A^\ast$ matrices have the same eigenvalues and complex conjugated eigenvectors, only matrix $A$ needs to be explicitly diagonalized so as to reduce the computational cost by a factor of two.
Moreover, unlike the quaternion structure \eqref{HCase1}, only $\{|J\mathbf{d}M\rangle\}$ (or equivalently $\{\mathcal{T}|J\mathbf{d}M\rangle\}$) are needed to expand the CI eigenvectors,
thereby reducing the computational cost by another factor of two.
As such, the combined use of spatial and time reversal symmetries leads to a reduction of the computation cost by a factor of $|G^\ast|$
(four for $C_2^\ast$ and $C_s^\ast$ and eight for $C_{2h}^\ast$).

Finally, the fermion irreps of  $C_{2v}^\ast$, $D_2^\ast$, and $D_{2h}^\ast$ belong to class (c). Such irreps are all two dimensional.
$\{|J\mathbf{d}M\rangle\}$ and $\{\mathcal{T}|J\mathbf{d}M\rangle\}$ can be arranged to different columns of the same irrep, i.e.,
$(|J\mathbf{d}M 1\rangle,|J\mathbf{d}M 2\rangle)$ with $|J\mathbf{d}M 2\rangle=\mathcal{T}|J\mathbf{d}M 1\rangle$ and $M>0$. The first-column functions $\{|J\mathbf{d}M 1\rangle\}$
are given in Tables \ref{TimeReversald22} and \ref{TimeReversalC2v} for $D_2^\ast$ and $C_{2v}^\ast$, respectively.
Such functions are characterized by the irreps of the spatial part of $|J\mathbf{d}M\rangle$ as well as the odd and even characters of $M-\frac{1}{2}$, totalling up to 8 cases.
The corresponding functions for $D_{2h}^\ast$ can be obtained from Table \ref{TimeReversald22} by replacing an irrep $X$ of $D_{2}$ with $X_g$ or $X_u$. With the so-constructed
Kramers-paired basis, the Hamiltonian matrix takes the following structure (cf. Eq. \eqref{reducedH})
\begin{equation}
(|J\mathbf{d}M1\rangle,\mathcal{T}|J\mathbf{d}M1\rangle)^\dagger H (|J\mathbf{d}M1\rangle,\mathcal{T}|J\mathbf{d}M1\rangle) =
\begin{bmatrix}
A & 0 \\
0 & A
\end{bmatrix},\label{HCase3}
\end{equation}
where matrix $A$ is real-valued and is independent of the column indices (cf. Eq. \eqref{TH1}. This structure
itself gives rises to a factor of 8 reduction of the computational cost.
%Just like class (b), only matrix $A$ is explicitly diagonalized,
%and one of $|J\mathbf{d}M1\rangle$ and $|J\mathbf{d}M2\rangle$ is involved to expand the eigenvectors.
%Furthermore, since the matrix $A$ is real, the CI vector is also real, further reducing the cost by a factor of two.
As such, the combined use of spatial and time reversal symmetries leads to a reduction of the computation cost by a factor of eight ($=|C_{2v}^\ast|=|D_2^\ast|$) for $C_{2v}^\ast$ and $D_2^\ast$ and sixteen ($=|D_{2h}^\ast|$) for $D_{2h}^\ast$,
corresponding to the fact that $C_{2v}^\ast$ and $D_2^\ast$ have only one but $D_{2h}^\ast$ has two fermion irreps.

\begin{table}
	\centering
	\caption{Symmetrized fermion functions of $C_1^\ast$ and $C_{i}^\ast$}
	\begin{threeparttable}
		\begin{tabular}{l|ccc}\toprule
			\multicolumn{1}{c|}{function}&\multicolumn{1}{c}{group}
			& irrep of $|J\mathbf{d}\rangle$
			& irrep  \\\toprule
			\multirow{4}{*}{$|J\mathbf{d}M\rangle$}
            &\multirow{1}{*}{$C_1^\ast$} & $A$    & $A_{1/2}$ \\
            & \\
            &\multirow{1}{*}{$C_i^\ast$} & $A_g$  & $A_{1/2,g}$ \\
			&                            & $A_u$  & $A_{1/2,u}$ \\ \midrule
		\end{tabular}
	\end{threeparttable}
	\label{TimeReversalCi}
\end{table}

\begin{table}
	\centering
	\caption{Symmetrized fermion functions of $C_{2}^\ast$. $M=q+1/2>0$}
	\begin{threeparttable}
		\begin{tabular}{c|ccc}\toprule
%		\multicolumn{1}{c|}{CSF}
		\multirow{2}{*}{function}
		& \multirow{2}{*}{irrep of $|J\mathbf{d}\rangle$}
		&\multicolumn{2}{c}{irrep}
		\\\cline{3-4}
		&
		& $q$ is odd
		& $q$ is even \\\toprule
		\multirow{2}{*}{$|J\mathbf{d}M\rangle$}
		& $A$  & $^1E_{1/2}$ & $^2E_{1/2}$ \\
		& $B$  & $^2E_{1/2}$ & $^1E_{1/2}$ \\ \midrule
		\multirow{2}{*}{$|J\mathbf{d}-M\rangle$}
		& $A$  & $^2E_{1/2}$ & $^1E_{1/2}$ \\
		& $B$  & $^1E_{1/2}$ & $^2E_{1/2}$ \\ \midrule
		\end{tabular}
	\end{threeparttable}
	\label{TimeReversalC2}
\end{table}

\begin{table}
	\centering
	\caption{Symmetrized fermion functions of $C_{s}^\ast$. $M=q+1/2>0$}
	\begin{threeparttable}
	\begin{tabular}{c|ccc}\toprule
		%		\multicolumn{1}{c|}{CSF}
		\multirow{2}{*}{function}
		& \multirow{2}{*}{irrep of $|J\mathbf{d}\rangle$}
		&\multicolumn{2}{c}{irrep}
		\\\cline{3-4}
		&
		& $q$ is odd
		& $q$ is even \\\toprule
			\multirow{2}{*}{$|J\mathbf{d}M\rangle$}
			& $A'$   & $^1E_{1/2}$ & $^2E_{1/2}$ \\
			& $A''$  & $^2E_{1/2}$ & $^1E_{1/2}$ \\ \midrule
			\multirow{2}{*}{$|J\mathbf{d}-M\rangle$}
			& $A'$   & $^2E_{1/2}$ & $^1E_{1/2}$ \\
			& $A''$  & $^1E_{1/2}$ & $^2E_{1/2}$ \\ \midrule
		\end{tabular}
	\end{threeparttable}
	\label{TimeReversalCs}
\end{table}

\begin{table}
	\centering
	\caption{Symmetrized fermion functions of $C_{2h}^\ast$. $M=q+1/2>0$}
	\begin{threeparttable}
		\begin{tabular}{c|ccc}\toprule
			%		\multicolumn{1}{c|}{CSF}
			\multirow{2}{*}{function}
			& \multirow{2}{*}{irrep of $|J\mathbf{d}\rangle$}
			&\multicolumn{2}{c}{irrep}
			\\\cline{3-4}
			&
			& $q$ is odd
			& $q$ is even \\\toprule
			\multirow{4}{*}{$|J\mathbf{d}M\rangle$}
			& $A_g$  & $^1E_{1/2,g}$ & $^2E_{1/2,g}$ \\
			& $B_g$  & $^2E_{1/2,g}$ & $^1E_{1/2,g}$ \\
			& $A_u$  & $^1E_{1/2,u}$ & $^2E_{1/2,u}$ \\
			& $B_u$  & $^2E_{1/2,u}$ & $^1E_{1/2,u}$ \\ \midrule
			\multirow{4}{*}{$|J\mathbf{d}-M\rangle$}
			& $A_g$  & $^2E_{1/2,g}$ & $^1E_{1/2,g}$ \\
			& $B_g$  & $^1E_{1/2,g}$ & $^2E_{1/2,g}$ \\
			& $A_u$  & $^2E_{1/2,u}$ & $^1E_{1/2,u}$ \\
			& $B_u$  & $^1E_{1/2,u}$ & $^2E_{1/2,u}$ \\ \midrule
		\end{tabular}
	\end{threeparttable}
	\label{TimeReversalC2h}
\end{table}

\begin{table}
	\centering
	\caption{The first column $|J\mathbf{d}M 1\rangle$ of the basis $(|J\mathbf{d}M 1\rangle,|J\mathbf{d}M 2\rangle)$ for the fermion irreps of $D_{2}^\ast$. $M=q+1/2>0$ }
	\begin{threeparttable}
		\begin{tabular}{c|rr}\toprule
			irrep of $|J\mathbf{d}\rangle$
			& \multicolumn{1}{c}{$q$ is odd}
			& \multicolumn{1}{c}{$q$ is even} \\ \toprule
			$A$   & $|J\mathbf{d}M\rangle$                & $|J\mathbf{d}-M\rangle$ \\
			$B_1$ & $\mathbbm{i}|J\mathbf{d}M\rangle$     & $ \mathbbm{i}|J\mathbf{d}-M\rangle$ \\
			$B_2$ & $|J\mathbf{d}-M\rangle$               & $|J\mathbf{d}M\rangle$ \\
			$B_3$ & $\mathbbm{i}|J\mathbf{d}-M\rangle$    & $\mathbbm{i}|J\mathbf{d}M\rangle$ \\\bottomrule
	\end{tabular}
	\end{threeparttable}
	\label{TimeReversald22}
\end{table}

\begin{table}
	\centering
	\caption{The first column $|J\mathbf{d}M 1\rangle$ of the basis $(|J\mathbf{d}M 1\rangle,|J\mathbf{d}M 2\rangle)$ for the fermion irreps of $C_{2v}^\ast$. $M=q+1/2>0$ }
	\begin{threeparttable}
		\begin{tabular}{c|rr}\toprule
			irrep of $|J\mathbf{d}\rangle$
			& \multicolumn{1}{c}{$q$ is odd}
			& \multicolumn{1}{c}{$q$ is even} \\ \toprule
			$A_1$ & $|J\mathbf{d}M\rangle$                & $|J\mathbf{d}-M\rangle$ \\
			$A_2$ & $\mathbbm{i}|J\mathbf{d}M\rangle$     & $ \mathbbm{i}|J\mathbf{d}-M\rangle$ \\
			$B_1$ & $|J\mathbf{d}-M\rangle$               & $|J\mathbf{d}M\rangle$ \\
			$B_2$ & $\mathbbm{i}|J\mathbf{d}-M\rangle$    & $\mathbbm{i}|J\mathbf{d}M\rangle$ \\\bottomrule
		\end{tabular}
	\end{threeparttable}
	\label{TimeReversalC2v}
\end{table}

\clearpage
\newpage

\bibliography{iCI}
\bibliographystyle{iopart-num}

\clearpage
\newpage

%For TOC only \\
%
%\begin{figure}
%\centering
%	\includegraphics[width=1.0\textwidth]{FigureSOiCI}
%\end{figure}

\end{document}